\documentclass[a4paper,11pt]{article}
\pdfoutput=1

\usepackage{jcappub} 

\usepackage[T1]{fontenc} 
\usepackage[
colorlinks=true, 
linkcolor=blue,
citecolor=blue,
urlcolor=blue
]{hyperref}

\title{$A'$ view of the sunrise: Boosting \mbox{helioscopes with angular information}}

\author[a]{Jonas Frerick}
\author[b]{Felix Kahlhoefer}
\author[a]{Kai Schmidt-Hoberg}

\affiliation[a]{Deutsches Elektronen-Synchrotron DESY,\\ Notkestr.~85, 22607 Hamburg, Germany}
\affiliation[b]{Institute for Theoretical Particle Physics (TTP), Karlsruhe Institute of Technology (KIT),\\ 76128
Karlsruhe, Germany}

\emailAdd{jonas.frerick@desy.de}
 \emailAdd{kahlhoefer@kit.edu}
\emailAdd{kai.schmidt-hoberg@desy.de}

\arxivnumber{DESY-22-165, TTP22-065}

\abstract{
The Sun may copiously produce hypothetical light particles such as axions or dark photons, a scenario which can be experimentally probed with so-called helioscopes. Here we investigate the impact of the angular and spectral distribution of solar dark photons on the sensitivity of such instruments. For the first time we evaluate this spectral and angular dependence of the dark photon flux over the whole mass range and apply this information to existing data from the Hinode Solar X-Ray Telescope. Specifically we use calibration images for a classical helioscope analysis as well as data from a solar eclipse providing sensitivity to exceptionally large oscillation lengths. We demonstrate that exploiting the signal features can boost the constraints by more than one order of magnitude in terms of the mixing parameter compared to a naive counting experiment. 
}


\begin{document}

\maketitle

\section{Introduction}
\label{sec:intro}

The evidence for dark matter (DM) in the universe makes us confident that there exists physics beyond the Standard Model (SM). Nevertheless, the lack of exciting signals in spite of huge experimental efforts suggests that this new physics may differ from common expectations and may for example manifest itself at much smaller masses and couplings than usually assumed. One of the most well-motivated ways to add new light degrees of freedom to the SM are so-called dark photons (DPs)~\cite{HOLDOM1986196,Fabbrichesi_2021}, i.e.\ (massive) vector bosons which interact with the SM via kinetic mixing. These dark photons are known to be viable DM candidates across a wide range of masses and predict a number of interesting observable effects even if they only constitute a small fraction of DM.

In this work we will consider the simple case that the dark photon has no gauge interactions with any other particles from the visible or dark sector, such that the model is fully characterised by the mixing parameter $\epsilon$ and the DP mass $m$. The most common approach to generate a mass for the DP is the St\"{u}ckelberg mechanism, which circumvents the strong exclusion limits on the case of a spontaneously broken $U(1)$ via a dark Higgs boson~\cite{Ahlers:2008qc,Redi:2022zkt}.

Constraints on DPs stretch from the GeV-regime that can be investigated in colliders or beam dump experiments \cite{Ferber:2022ewf, Bauer:2018onh,Graham:2021ggy} to the realm of so-called fuzzy DM \cite{Hu:2000ke,Kovetz:2018zes} at around $10^{-21}$eV, which is extremely hard to probe directly. The strength of these constraints depends on whether the DP is considered as a DM candidate or as a more general extension of the SM. In the former case, it is especially interesting to consider those regions of parameter space that can be explored using haloscopes~\cite{Suzuki:2015sza,Baryakhtar:2018doz,An:2022hhb,Cervantes:2022gtv}, i.e.\ experiments dedicated to detecting the presence of DM in the Milky Way halo. In the latter case, searches for DPs are either indirect (exploiting for example virtual effects) or rely on local production mechanisms, leading to substantially more freedom in the parameter space.

A local DP source of particular importance is the Sun, which may produce large numbers of DPs with mass smaller than or comparable to its temperature (or, more precisely, its plasma frequency)~\cite{2008,2015,An_2013,An:2013yua,Redondo_2013}. Due to plasma effects, which can resonantly enhance or strongly suppress DP production, the production rate depends sensitively on the DP mass. The resulting ``dark'' luminosity of the Sun can be used to constrain DP models by requiring that the visible luminosity remains the dominant form of cooling. This so-called cooling argument can also be applied to other astrophysical environments, such as horizontal branch stars, red giants \cite{Redondo_2013} or neutron stars \cite{Hong:2020bxo}. Furthermore, DP plasma effects play a crucial role in the effects that DPs can have on the early universe~\cite{Caputo:2020rnx,Mirizzi:2009iz,Caputo:2020bdy,Bondarenko:2020moh,Garcia:2020qrp,Witte:2020rvb} where e.g.\ resonant photon-DP conversion can distort the CMB black body spectrum. Furthermore, plasma effects might also be relevant for black hole superradiance considerations~\cite{Caputo:2021efm} where a local electron density can quench the superradiant instability. 

In addition to these indirect constraints, we can hope to directly observe the flux of DPs produced in the Sun. The strategy is analogous to solar axion searches \cite{Sikivie:1983ip,vanBibber:1988pg}, i.e.\ it relies on the oscillation of DPs into SM photons over a finite distance (called oscillation length). In fact axion helioscopes like CAST~\cite{CAST:2002aym} can be directly repurposed to search for DPs~\cite{2008}. Since DPs do not require a magnetic field to oscillate into SM photons, it is also possible to construct simpler dedicated DP-only helioscopes, such as SHIPS~\cite{2015ships}. These searches constrain a wide range of DP masses from several keV down to $\mu$eV. Even stronger constraints stem from searches for the absorption of the longitudinal DP component in direct detection experiments \cite{An_2013,An:2013yua,An:2020bxd,XENON:2021nad}.

In this work we point out that constraints on DPs from helioscopes may be improved considerably by taking into account the spectral and angular distribution of DPs, which has not previously been considered in the context of DP searches (although it has been studied in the context of axion searches~\cite{CAST:2017uph}\footnote{In fact axions are always predominantly produced in the centre of the Sun.}). To illustrate our argument, we analyse the publicly available data of the Hinode X-Ray Telescope (XRT) \cite{2007SoPh..243...63G}, which offers excellent angular resolution.  We show that it is possible to infer spectral information from this instrument even though it was not originally built for this purpose. We then use the relatively long exposure of so-called ``darks'', calibration images of the Sun with a closed telescope, to demonstrate the improvement of the constraints with respect to the naive analysis of just counting events without spectral or angular information like in previous helioscope searches. We find that thanks to this improvement, Hinode XRT can almost match the sensitivity to transverse DPs of purpose-built helioscopes.

We also analyse the data from Hinode XRT from an observed solar eclipse. This observation combines the main advantage of helioscopes, namely the large DP flux produced in the Sun, with an advantage of so-called light-shining-through-the-wall (LSW) experiments \cite{Okun:1982xi,Anselm:1987vj,VanBibber:1987rq}, namely a much longer oscillation length. Indeed, in a ``light shining through the Moon'' experiment, the oscillation length can be as long as the distance from the Moon to the Earth. We show that this advantage can compensate for the limited exposure, leading to a substantially improved sensitivity to DP masses below the meV scale compared to conventional helioscopes. Given that there exist multiple x-ray satellites that constantly track(ed) the solar activity~\cite{doi:10.1126/science.258.5082.618,2007SoPh..243...63G,article}, we expect that even stronger constraints can be obtained by combining data from several instruments and multiple solar eclipses.

The remainder of this paper is structured as follows. In section \ref{sec:prod} we will briefly discuss the solar production of DPs with emphasis on the angular distribution in different production regimes. 
In section~\ref{sec:analysis} we present a collection of telescopes of interest, identify the requirements for our analysis and select a useful data set. We then summarise our analysis strategy, emphasising the relevance of proper calibration, background subtraction, and knowledge of the angular and spectral distribution to place more stringent limits. We subsequently contextualise our results and discuss if and how experiments of this type can be made competitive with the leading constraints in the mass range of interest. Finally, we conclude in section \ref{sec:conc}. Additional technical details of our analysis are summarised in the appendix. Throughout this paper, we work with $\hbar=c=1$.

\section{Solar production of DPs and their detection}
\label{sec:prod}

In this section, we briefly review the main properties of DPs in vacuum and then generalise to homogeneous plasma environments. We then discuss how to use this knowledge to calculate the flux of solar DPs and their angular dependence.

\subsection{DP oscillations in the vacuum and in homogeneous plasmas}
\label{sec:vacuum}
Let us start with a brief discussion of the vacuum physics of a massive DP $A^\prime_\mu$. The most general renormalisable and gauge-invariant Lagrangian for $A^\prime_\mu$ and the photon $A_\mu$ in vacuum can be written as
\begin{equation}
    \mathcal{L}_\mathrm{vac}= -\frac{1}{4} F^{\mu\nu}F_{\mu\nu}-\frac{1}{4} F^{\prime\mu\nu}F^\prime_{\mu\nu}-\frac{\epsilon}{2}F^{\mu\nu}F^\prime_{\mu\nu}+j_\mu A^\mu+\frac{m^2}{2}A^{\prime\mu}A^\prime_\mu\; , \label{eq:vaclag}
\end{equation}
where the off-diagonal kinetic term couples the dark and SM sectors, with $F^{\mu\nu}$ ($F^{\prime\mu\nu}$) being the electromagnetic (dark) field strength. A redefinition of the fields can yield the usual canonical (diagonal) kinetic structure, leading to either off-diagonal mass or interaction terms. We emphasise that in vacuum there are only two new parameters, namely the DP mass $m$ and the mixing angle $\epsilon$.

The kinetic mixing term allows for DP-photon oscillations in complete analogy to the case of neutrinos \cite{Smaldone:2021mii,Wolfenstein:1977ue} and will therefore allow for DP production from photons. In the case of small mixing, $\epsilon\ll1$, we have
\begin{equation}
    P(\text{DP}\leftrightarrow\gamma)=(2\epsilon)^2\sin^2\left(\frac{m^2L}{4\omega}\right)\; ,\label{eq:oscvac}
\end{equation}
for the relativistic vacuum oscillation probability, where $L$ denotes the distance that the DP/photon travels in the vacuum and $\omega$ stands for the particle's energy. For $m^2 L \gg \omega$, we expect the probability to oscillate so rapidly that it effectively averages out to $\frac{1}{2}$ and the physics on macroscopic experimental scales becomes independent of the exact energy and length dependence. This observation implies in particular that we can use eq.~\eqref{eq:oscvac} even when the assumption of fully relativistic DPs breaks down. Conversely, for $m^2 L \ll \omega$ we can use the small-angle approximation to show that the oscillation probability becomes proportional to $m^4$. In other words, any experiment that aims to detect DPs by relying on their conversion into photons will rapidly lose sensitivity for DP masses $m < \sqrt{\omega / L}$. This applies in particular to terrestrial helioscopes searching for DPs produced in the interior of the Sun, see section~\ref{sec:disc} and table~\ref{tab:osc} for further discussion.

While the physics of DPs in vacuum is remarkably simple, the environment can strongly impact the kinetic mixing portal. For a first understanding, let us review the simplified calculation from Ref.~\cite{Knapen:2017xzo}. Inside a medium such as the solar plasma, the photon acquires an effective mass $m_\gamma$~\cite{Popov_1991}, which we can approximate by the plasma frequency $\omega_p$:
\begin{equation}
    m_\gamma\approx\omega_p=\sqrt{\frac{4\pi\alpha n_e}{m_e}}\; 
\end{equation}
with the electron mass $m_e$ and density $n_e$. 
We now consider the following Lagrangian inside the Sun:
\begin{equation}
    \mathcal{L}= -\frac{1}{4} F^{\mu\nu}F_{\mu\nu}-\frac{1}{4} F^{\prime\mu\nu}F^\prime_{\mu\nu}-\frac{\epsilon}{2}F^{\mu\nu}F^\prime_{\mu\nu}+j_\mu A^\mu+\frac{m^2}{2}A^{\prime\mu}A^\prime_\mu+\frac{m_\gamma^2}{2} A^\mu A_\mu\; .
\end{equation}
Note that there is an implicit radial dependence of the photon mass $m_\gamma=m_\gamma(\mathbf{r})=m_\gamma(r)$ where the last equality follows under the assumption of spherical symmetry.
The introduction of an additional mass scale leads to two extremal limits. For the case of DPs much heavier than the photon plasma mass, $m \gg m_\gamma$, we effectively recover the vacuum scenario such that we can ignore plasma effects. The opposite scenario of a dominant plasma mass, $m \ll m_\gamma$, will lead to a gradual decoupling of the DP consistent with the well-known case of a completely decoupled massless DP (with an otherwise empty dark sector).\footnote{Note that this appears to be in contradiction with the statement that the massless DP and the massive DP in the massless limit are not the same \cite{Fabbrichesi_2021}, i.e.\  $``m=0"\neq ``m\rightarrow0"$. While this is true in a perfect vacuum where the DP mass will always set the largest mass scale in the system of eq.~\eqref{eq:vaclag}, realistic environments will always lead to matter effects which set an additional mass scale that can effectively suppress the kinetic mixing. Thus, the following calculation will not be in contradiction with the conceptual difference between massive and massless DPs.} 

To explore these limiting cases, let us restore the canonical kinetic terms up to first order in the mixing parameter $\epsilon$,
\begin{gather}
    A_\mu\rightarrow A_\mu \nonumber\\
    A^\prime_\mu\rightarrow A^\prime_\mu -\epsilon A_\mu\nonumber\\ 
     \Rightarrow\mathcal{L}= -\frac{1}{4} F^{\mu\nu}F_{\mu\nu}-\frac{1}{4}F^{\prime\mu\nu}F^\prime_{\mu\nu}+j_\mu A^ \mu+\frac{m^2}{2}A^{\prime\mu}A^\prime_\mu-\epsilon m^2A^\mu A^\prime_\mu+\frac{m_\gamma^2}{2} A^\mu A_\mu\; .
     \label{eq:mixlag}
\end{gather}
This form of the Lagrangian corresponds to the 
interaction eigenbasis, in which the DP is completely decoupled from the SM fermions and can only oscillate into the photon (and vice versa) via the mass mixing term. 
We will mostly work with this basis in the following. Nevertheless, to improve our understanding of the oscillation probability, let us briefly also investigate the propagation eigenbasis, where the mass terms are diagonal. To achieve this diagonalisation while simultaneously keeping the canonical kinetic terms intact we perform an orthogonal rotation, which introduces an effective coupling of the DP to the electromagnetic current
\begin{equation}
j_\mu A^\mu\rightarrow j_\mu(A^\mu+\epsilon_{\text{eff}} A^{\prime\mu}) 
\end{equation}
with
\begin{equation}
   \epsilon_{\text{eff}}=\epsilon \frac{m^2}{|m_\gamma^2-m^2|} \approx \begin{cases} \epsilon \, , & m^2\gg m_\gamma^2 \\ \epsilon \frac{m^2}{m_\gamma^2} \, , & m^2\ll m_\gamma^2\end{cases}\; .
\end{equation}
Again we recover the decoupling of the DP in the limit of dominant plasma mass.

The intermediate regime where $m\approx m_\gamma$ allows for resonant conversions between photons and DPs and therefore needs to be treated with more care. In this resonant regime, we will find that the spectral and the angular part of the DP flux factorise, simplifying the calculation significantly compared to the two limiting regimes. We will use a more careful treatment of plasma effects to study this case in the next subsection.

\subsection{Solar DP physics}

From now on, we will follow closely Refs.~\cite{2008,2015}, which provides a detailed description of the solar DP production covering mostly the transverse components. Further work on solar DP production can be found in Refs.~\cite{An_2013,Redondo_2013} with a focus on the longitudinal component, which is not of interest in this work because it cannot be detected in classical helioscope setups (see section~\ref{sec:oscillation}). We will give a detailed description of the solar DP production based on input from the solar standard model \cite{Bahcall:2004pz,Caffau:2010qc,Magg:2022rxb}. For the first time, we give estimates for the angular distribution of DPs over the whole mass spectrum.
For the remainder of this section, we will exclusively work in the interaction eigenbasis of eq.~\eqref{eq:mixlag}.

\begin{figure}
    \centering
    \includegraphics[width=0.8\linewidth,clip,trim=0 50 0 50]{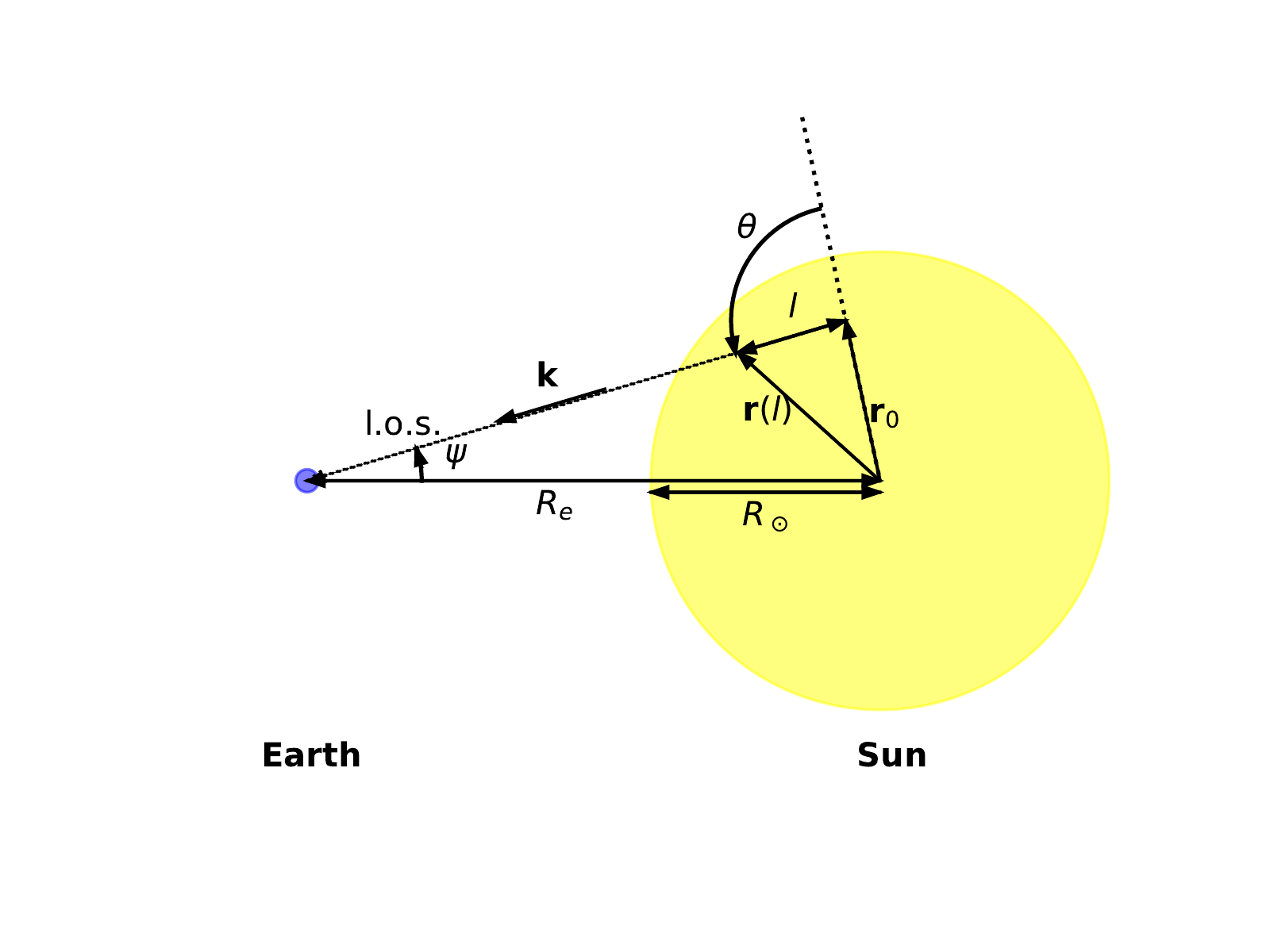}
    \caption{In this sketch of the Sun-Earth system (adapted from Ref.~\cite{2015}), we can define all the geometric quantities necessary to calculate the solar DP flux. As we work in the approximation of an isotropic solar model, the angular flux can at most depend on the angle $\psi$ under which we observe the Sun.}
    \label{fig:sketch}
\end{figure}

Let us start with the production rate of DPs, which can be written as~\cite{2008}
\begin{equation}
    \frac{\mathrm{d}N}{\mathrm{d}V\mathrm{d}t}(\mathbf{r}_0)=2\int \frac{\mathrm{d}^3k}{(2\pi)^3} \Gamma_\mathrm{prod}(\omega,\mathbf{r}_0) P(\omega,\mathbf{r}(l))\; . \label{eq:prodnum}
\end{equation}
Here the factor of $2$ accounts for the two transverse helicities of the DP as we ignore the longitudinal component. The local production rate of photons follows from local thermodynamic equilibrium and is given by
\begin{equation}
    \Gamma_\mathrm{prod}(\omega,\mathbf{r}_0)=\frac{\Gamma (\omega,\mathbf{r}_0)}{e^{\omega/T(\mathbf{r}_0)}-1}\; ,\label{eq:prod}
\end{equation}
where $\Gamma=\Gamma_\mathrm{abs}-\Gamma_\mathrm{prod}\,$, i.e.\ we define the so-called absorption coefficient $\Gamma$ as the difference of absorption and production rates.
Furthermore, we have used detailed balance to obtain the relation $\Gamma_\mathrm{prod}=e^{-\omega/T}\Gamma_\mathrm{abs}$, where $T$ is the (position dependent) temperature. To obtain the DP production rate the local production rate of photons must be multiplied with the probability $P(\omega,\mathbf{r}(l))$ that the photon oscillates into a DP, which depends on the trajectory of the photon $\mathbf{r}(l)=\mathbf{r}_0+l\hat{\mathbf{k}}$ where $l$ denotes the distance the photon travels and $\hat{\mathbf{k}}$ denotes its direction, see figure~\ref{fig:sketch}. 
Finally, the total DP production is given by an integral over the photon phase space, i.e.\ all possible momenta. 

Using the on-shell relation $\omega^2=k^2+m^2$ and the rotational symmetry of the momentum in eq.~\eqref{eq:prodnum} we find 
\begin{align}
    \frac{\mathrm{d}N}{\mathrm{d}V\mathrm{d}t}=&\frac{4\pi}{(2\pi)^3}\int \mathrm{d}\cos\theta_k k^2 \mathrm{d}k \frac{\Gamma (\omega,\mathbf{r}_0)}{e^{\omega/T(\mathbf{r}_0)}-1}P(\omega,\mathbf{r}(l))\nonumber\\
    \Rightarrow \frac{1}{4\pi R_e^2}\frac{\mathrm{d}N}{\mathrm{d}V\mathrm{d}t \mathrm{d}\omega }=&\frac{\mathrm{d}\Phi}{\mathrm{d}V\mathrm{d}\omega}=\frac{\omega\sqrt{\omega^2-m^2}}{8\pi^3R_e^2}\int \mathrm{d}\cos\theta_k \frac{\Gamma (\omega,\mathbf{r}_0)}{e^{\omega/T(\mathbf{r}_0)}-1}P(\omega,\mathbf{r}(l))\; .
\end{align}
Here, we have defined the flux $\Phi$ at the position of Earth, i.e.\ at a distance of $R_{e}$. One can now obtain the total spectral flux of the Sun by integrating over the Sun's volume:
\begin{align}
    \frac{\mathrm{d}\Phi}{\mathrm{d}\omega}=&\frac{\omega\sqrt{\omega^2-m^2}}{8\pi^3 R_e^2 }\int_{\text{sun}} \mathrm{d}V \int \mathrm{d}\cos\theta_k \frac{\Gamma (\omega,r)}{e^{\omega/T(r)}-1}P(\omega,\mathbf{r}(l))\nonumber\\
    =&\frac{\omega\sqrt{\omega^2-m^2}}{2\pi^2 R_e^2}\int_0^{R_\odot} r^2\mathrm{d}r \int_{-1}^1 \mathrm{d}\cos\theta \frac{\Gamma (\omega,r)}{e^{\omega/T(r)}-1}P(\omega,r,\theta)\; ,\label{eq:total}
\end{align}
where $R_\odot$ denotes the solar radius and we have used that the probability $P$ can only depend on the radial position $r$ and on the angle $\theta$ between the position vector $\mathbf{r}$ and the photon direction $\mathbf{k}$ (i.e.\ $\mathbf{r}\cdot \mathbf{k}=rk\cos\theta$) so that we can integrate over all remaining angles.

In order to calculate the angular distribution of DPs, we need to consider the DP production along a chosen line of sight (l.o.s.) as shown in figure~\ref{fig:sketch}. We can parametrise this l.o.s. by the radius $r$ and the point of closest distance to the core $r_\text{min}$. This yields
\begin{align}
    \frac{\mathrm{d}\Phi}{\mathrm{d}\omega \mathrm{d}\Omega}=& \frac{\omega\sqrt{\omega^2-m^2}}{4\pi^3}\int_{\text{l.o.s.}} \mathrm{d}s \frac{\Gamma (\omega,r)}{e^{\omega/T(r)}-1}P(\omega,\mathbf{r}(l)) \nonumber \\
    =& \frac{\omega\sqrt{\omega^2-m^2}}{4\pi^3}\int_{r_\text{min}}^{R_\odot} \frac{2r \mathrm{d}r}{\sqrt{r^2-r_\text{min}^2}} \frac{\Gamma (\omega,r)}{e^{\omega/T(r)}-1}P(\omega,r,\theta)\; .
\label{eq:truangu}
\end{align}
Following Ref.~\cite{2015} we assume that the oscillation length of the DP is always much smaller than the scale of significant changes in the plasma, i.e.\ we take the plasma to be locally homogeneous. For weak mixing the in-medium oscillations can then simply be described by~\cite{2008,Redondo_2013,An_2013,2015}
\begin{equation}
    P(\gamma \rightarrow\text{DP}) =\frac{\epsilon^2m^4 }{\left(m_\gamma^2(r)-m^2\right)^2+\left(\omega\Gamma(\omega,r)\right)^2}\; ,\label{eq:matosc}
\end{equation}
where $\Gamma$ again denotes the absorption coefficient. The plasma mass and the absorption coefficient are important for position and strength of a potential resonance in the conversion probability. We note that the former is given by the real and the latter by the imaginary part of the transverse component of the photon self-energy \cite{PhysRevD.28.2007,2008} which enables us to relate the microphysics of the oscillation to the macroscopic plasma properties like its electron density. 

If the DP mass is such that the condition $m \approx m_\gamma(r)$ can be fulfilled for some specific radius $r = r_\ast$, DP production will be dominated by a thin shell around $r_\ast$. As shown in detail in appendix~\ref{sec:app-a}, we find that in this resonant regime the differential flux factorises into a spectral and an angular part
\begin{align}
 \frac{\mathrm{d}\Phi_\text{res}}{\mathrm{d}\omega \mathrm{d}\Omega}\approx \frac{\mathrm{d}\Phi}{\mathrm{d}\omega} \frac{\mathrm{d}X_\text{res}}{\mathrm{d}\Omega}\; ,
    \label{eq:factorisation}
    \end{align}
with $\mathrm{d}\Phi/\mathrm{d}\omega$ given by eq.~\eqref{eq:total} and 
\begin{align}
    \frac{\mathrm{d}X_\text{res}}{\mathrm{d}\Omega}=\frac{1}{2\pi}\frac{R_e}{r_\ast} \frac{1}{\sqrt{\psi_\ast^2-\psi^2}}\Theta(r_\ast-r_\text{min})\; .
    \end{align}
Here we follow the notation from Ref.~\cite{2015}, i.e.\ we introduce the observation angle $\psi=\frac{r_\text{min}}{R_{e}}$ (see figure~\ref{fig:sketch}) and the resonance angle $\psi_\ast=\psi\big|_{m=m_\gamma}$. Our final result however differs slightly from Ref.~\cite{2015}, in particular in that it is correctly normalised, i.e.\ $\int \mathrm{d}\Omega \tfrac{\mathrm{d}X}{\mathrm{d}\Omega} = 1$.

The resonance condition can only be met for a small range of DP masses around 5--300$\,$eV\footnote{For $m < 5\,\mathrm{eV}$ one would in principle expect resonant conversion in the outer regions of the Sun, where the assumption of spherical symmetry breaks down. However, the exponential suppression in eq.~\eqref{eq:prod} makes the resonant contribution negligible.}, while helioscopes are in principle sensitive to the much wider range from around $10^{-6}$--$10^3\,$eV. In this non-resonant case, there is no factorisation of angular and energy dependence because the absorption term depends inseparably on position and energy. Since a full numerical calculation of the differential flux is computationally expensive, it is desirable to obtain at least approximate expressions for the angular distribution also in this case. To the best of our knowledge, these expressions have not been presented previously, even though they can be derived based on relatively simple arguments.

First of all, we should note that there a two distinct regimes: suppressed ($m<m_{\gamma,\text{min}}\sim 5\,\mathrm{eV}$) and unsuppressed ($m>m_{\gamma,\text{max}}\sim 300\,\mathrm{eV}$) with only a weak energy dependence. 
Let us begin with the unsuppressed case, which we label with subscript vac, as it is similar to the vacuum scenario. For fixed values of $\epsilon$ and $m$ we find
\begin{align}
 P_\text{vac}(\omega,r,\theta)\approx &\frac{\epsilon^2 m^4}{\left(m_\gamma^2(r)-m^2\right)^2+\left(\omega \Gamma(\omega,r)\right)^2}
    \xrightarrow{m\gg m_\gamma} \text{const} \nonumber\\
    \Rightarrow \frac{\mathrm{d}\Phi_\text{vac}}{\mathrm{d}\omega \mathrm{d}\Omega}\propto& \int_{r_\text{min}}^\infty \frac{2r \mathrm{d}r}{\sqrt{r^2-r_\text{min}^2}} \frac{n_e(r)}{e^{\omega/T(r)}-1}\; ,
\end{align}
and similarly for the suppressed case
\begin{align}
 P_\text{sup}(\omega,r,\theta)&\xrightarrow{m \ll m_\gamma} \frac{\text{const}}{m_\gamma^4(r)}\nonumber\\
    \Rightarrow \frac{\mathrm{d}\Phi_\text{sup}}{\mathrm{d}\omega \mathrm{d}\Omega}\propto &\int_{r_\text{min}}^\infty \frac{2r \mathrm{d}r}{\sqrt{r^2-r_\text{min}^2}} \frac{n_e(r)}{e^{\omega/T(r)}-1}m_\gamma^{-4}(r)\; .
    \end{align}
Here we made the crucial assumption that the absorption coefficient receives its dominant contribution from the electron density, which should at least track the general trend of the radial dependence of $\Gamma$. The angular dependence then stems directly from the integration limit $r_\text{min}$.

To obtain the angular distribution of the DP signal for a given telescope,
we can convolute the flux with the energy-dependent efficiency $Q(\omega)$ of the instrument under consideration, leading to
\begin{equation}
    \frac{\mathrm{d}X_\text{vac/sup}}{\mathrm{d}\Omega}\propto \int_{E_\text{min}}^{E_\text{max}} \mathrm{d}\omega\  Q(\omega)\frac{\mathrm{d}\Phi_\text{vac/sup}}{\mathrm{d}\omega \mathrm{d}\Omega}\; .\label{eq:weight}
\end{equation}
This approximation is accurate if the range of integration $[E_\text{min},E_\text{max}]$ is small enough that the DP production does not change significantly.
Using appropriate choices of $E_\text{min/max}$, eq.~\eqref{eq:weight} can be used both for the whole spectrum or for individual energy bins. 
For temperature, density, and plasma mass we can use data shown in Ref.~\cite{2008} taken from Ref.~\cite{Bahcall:2004pz}. 
As we can fix the normalisation by hand, this is all we need to calculate a detector-dependent angular distribution.
In conclusion, we can write
\begin{equation}
    \frac{\mathrm{d}\Phi_\text{res/vac/sup}}{\mathrm{d}\omega \mathrm{d}\Omega}\approx \frac{\mathrm{d}\Phi}{\mathrm{d}\omega} \frac{\mathrm{d}X_\text{res/vac/sup}}{\mathrm{d}\Omega}\; ,\label{eq:fact}
\end{equation}
with the different contributions as derived above. This factorisation enables us to make use of the predictions for $\frac{\mathrm{d}\Phi}{\mathrm{d}\omega}$ from Ref.~\cite{2015}.

\begin{figure}[t]
    \centering
    \includegraphics[width=0.8\columnwidth]{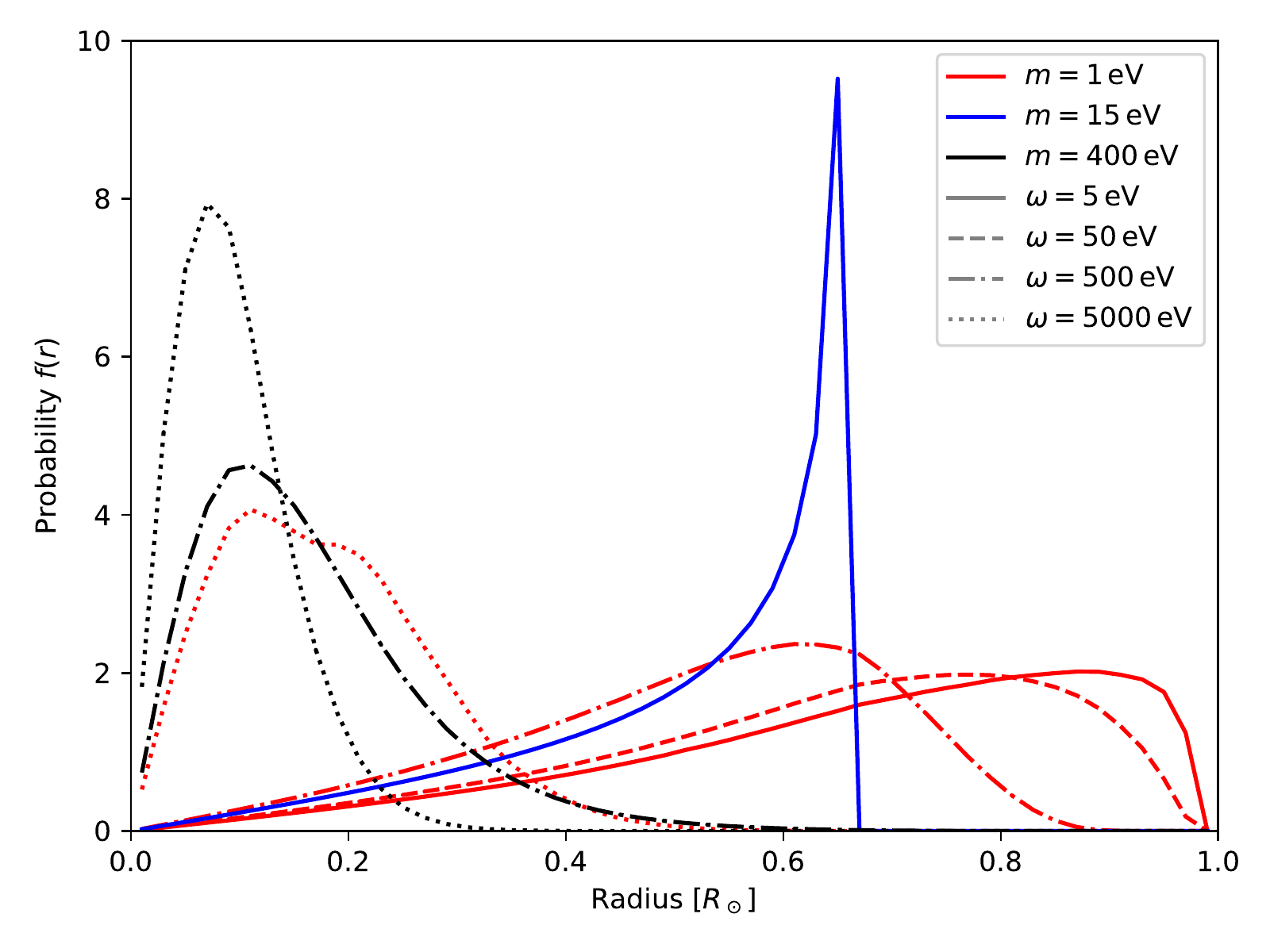}
    \caption{Plot of the different angular shapes of the DP signals for the suppressed, resonant and unsuppressed cases. $f(r)$ denotes the probability density of the DP distribution. We have normalised the distribution to the dimensionless radius $r/R_\odot$. We note that the resonant case is not energy dependent and the two remaining regimes are (approximately) mass independent except for the kinematic threshold of $m<\omega$. 
    }
    \label{fig:shapes}
\end{figure}

Before considering specific instruments with explicit efficiency functions in section~\ref{sec:analysis}, let us briefly look at the predictions without weighted averaging, i.e.\ for specific DP energies. For this purpose we define
\begin{equation}
    f(r)\equiv 2\pi \frac{r}{R_e} \frac{\mathrm{d}X}{\mathrm{d}\Omega}\frac{\mathrm{d}\psi}{\mathrm{d}r}=\frac{r}{R_e}\frac{\mathrm{d}X}{\mathrm{d}r},\label{eq:defprob}
\end{equation}
which gives the probability of a DP production event to occur in a ring of radius $r$ and width $\mathrm{d}r$ as $f(r)\mathrm{d}r$. We show this probability distribution in figure~\ref{fig:shapes}. As expected, the resonant regime exhibits a strong peak at $r = r_\ast$ where $m_\gamma(r_\ast)=m$. The position of this peak is energy independent as long as resonant production is kinematically allowed. In the suppressed and vacuum-like regime, the radial distribution is broader and energy dependent. The vacuum-like case has its peak close to the solar center, because it benefits from the high densities implying high photon production. The suppressed case experiences the same enhancement but also an additional suppression due to the higher plasma mass, which pushes the maximum further outwards, especially for small DP energies. For highly-energetic particles, the suppressed DP production will still come from the solar centre due to the exponential suppression of the photon production in colder regions of the Sun (see eq.~\eqref{eq:prod}).

\subsection{DP oscillation and detection}
\label{sec:oscillation}

Once a DP with transverse polarisation has been produced in the Sun, it can in principle oscillate back into a visible, i.e.\ interacting, photon. However, as long as the DP travels through the solar plasma, it will constantly be projected back onto the local propagation eigenstate, because the interacting component quickly gets absorbed. In fact, as the plasma density changes slowly and the absorption process is still efficient even in the outer regions of the Sun the DP will approach the \emph{vacuum propagation eigenstate}~\cite{2008}. Therefore, once the DP leaves the Sun and enters the vacuum\footnote{Vacuum here means that the mean free path of photons is large compared to the oscillation length $\lambda_\text{osc} = \omega / m^2$.} it cannot oscillate into visible photons irrespective of the distance it travels.

However, as soon as this propagation eigenstate enters dense matter (such as the shutter of a telescope or the Moon), the interaction component with electromagnetic interactions will be absorbed, such that one is left with the \emph{sterile interaction eigenstate}. If this state now rapidly transitions from dense matter to vacuum, it will no longer be in a local propagation eigenstate, such that the oscillation probability can build up. 

Contrary to transverse DPs, longitudinal DPs can only be detected for non-vanishing matter effects due to the transverse nature of the photon (in vacuum). As a result it is not possible for DPs with longitudinal polarisation to oscillate into visible photons in the usual helioscope setup. Nevertheless, the longitudinal mode can still be absorbed in a dense medium, which enables direct detection experiments to set constraints on longitudinal DPs.
For DP masses smaller than 1 eV, the longitudinal component is produced in larger numbers than the transverse one, leading to very impressive constraints from DM direct detection experiments \cite{An:2020bxd} (see also section~\ref{sec:disc}). The methods discussed in the following can in principle be applied also to these kinds of experiments provided a sufficient angular resolution can be achieved. However, we will focus on helioscopes in the remainder of this work, where data sets with excellent angular resolution are already available.

\section{Analysis}
\label{sec:analysis}

Having calculated the DP flux and oscillation probability, we can now turn to the question which telescope is best suited to search for the predicted signal. In the following we will identify the Hinode XRT as the best option to obtain data sets with good angular resolution. We will then proceed to analyse two different data sets: one with long exposure from calibration images with a shut telescope (so-called darks with an oscillation baseline of around $1\,$m) and one corresponding to a very long oscillation baseline of around $380000\,$km from the observation of a solar eclipse.

\subsection{The Hinode solar X-Ray Telescope} 

In order to set the most stringent constraints, it is clear that the time of solar observation should be as long as possible. For our purposes there are in fact two reasons: Firstly, long observation increases the chance to witness a solar eclipse. Secondly, it gives more statistics for the analysis of darks. Fortunately, there are a number of telescopes dedicated to solar observations. Out of these, we are primarily interested in x-ray telescopes, because the lunar and terrestrial x-ray albedos (and hence the probability to reflect x-rays from various sources) are extremely small \cite{10.1111/j.1365-2966.2008.12918.x}, ensuring that we operate in a region of minimal background. Moreover, compared to telescopes operating at lower frequencies, x-ray telescopes possess greater sensitivity to higher DP masses, which renders them more interesting to us.

With this collection of requirements, we are left with a small list of telescopes, including YOHKOH \cite{doi:10.1126/science.258.5082.618}, Hinode \cite{2007SoPh..243...63G} and the GOES series (see e.g.\ Ref.~\cite{article}). All these satellites carried/carry an x-ray telescope with them, and the relevant data is publicly available. A key difference between them is the height of the orbit, which varies between a few hundred kilometres for the Sun-synchronous orbit of Hinode and around 36,000 km for the geostationary orbit of GOES-12. These differences matter because the height of the orbit determines the dominant background sources (see below for a more detailed discussion). 

\begin{figure}[t]
    \centering
    \includegraphics[width=0.47\columnwidth]{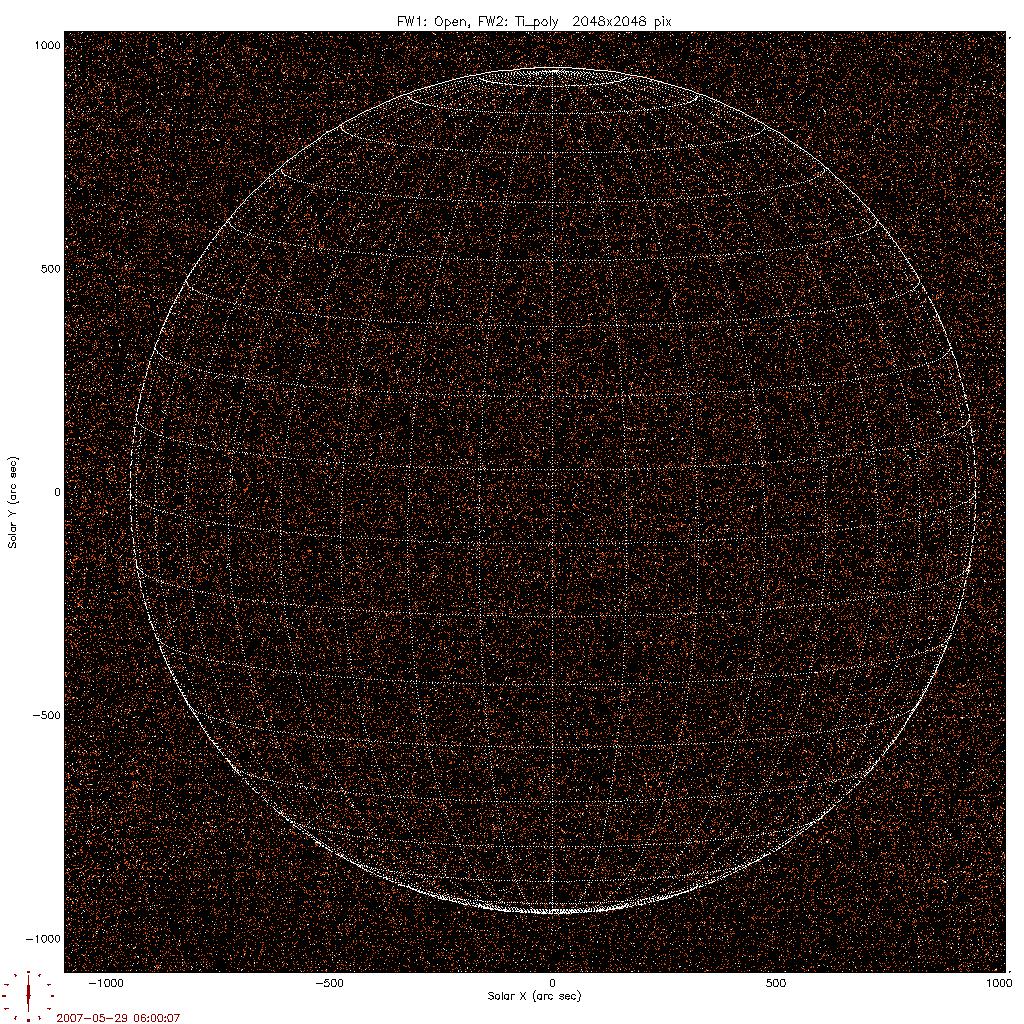}
    \includegraphics[width=0.47\columnwidth]{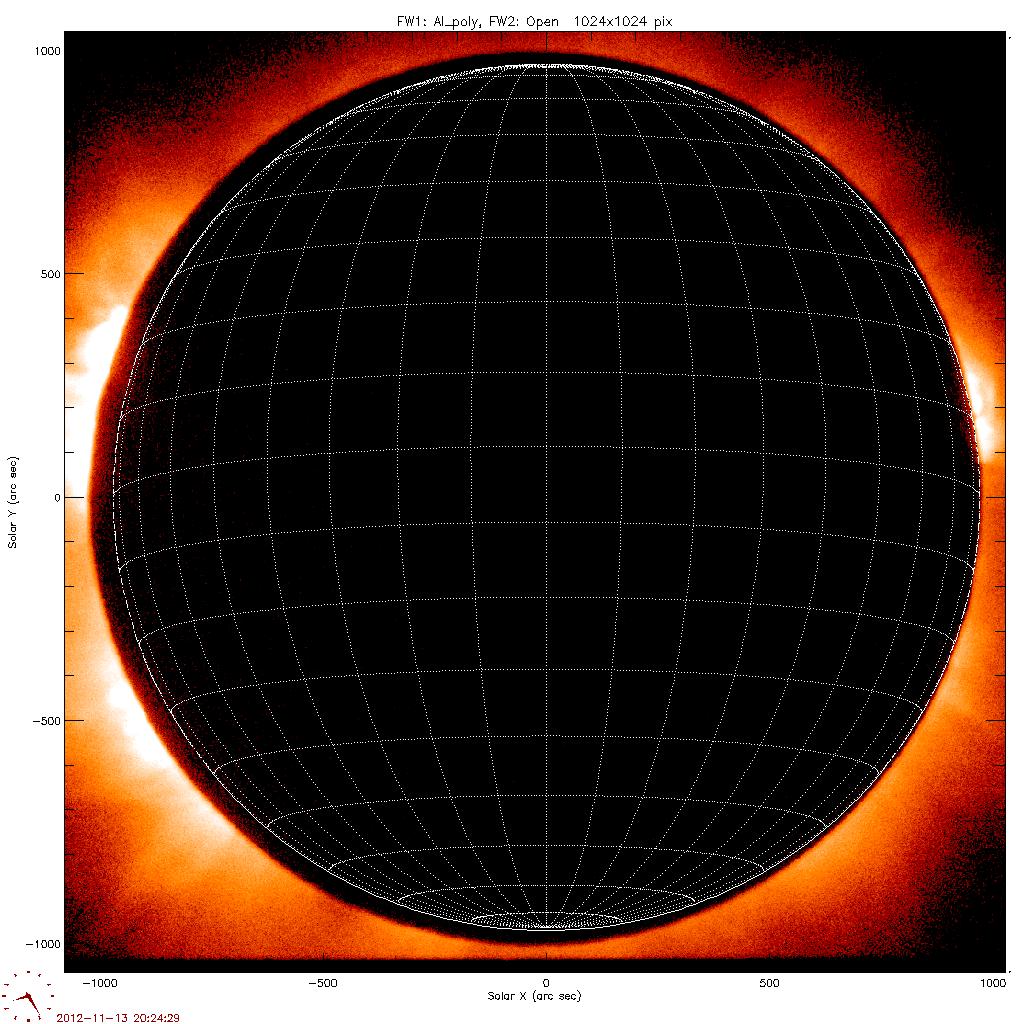}
    \caption{Images from \url{http://sdc.uio.no/search/form}. Left: Typical calibration image (dark) with $2048\times2048$ pixels. Right: Solar eclipse from 2012 with $1024\times1024$ pixels.}
    \label{fig:eclipses}
\end{figure}

For reasons of exposure, availability, and documentation, we have decided to analyse the Hinode XRT data \cite{Kosugi2007,2007SoPh..243...63G,2008SoPh..249..263K,Narukage_2011,Narukage_2013,Matsuzaki2007}.\footnote{All images are available at \url{http://sdc.uio.no/search/form} which also offers very helpful search criteria.} Figure~\ref{fig:eclipses} shows two images representing the two different data sets that we consider: a calibration image on the left and a solar eclipse on the right. The original purpose of the calibration images was mainly to collect statistics for a proper background subtraction, including time-varying effects. Nevertheless, they can be analysed in complete analogy to Earth-based helioscopes with an oscillation baseline $L$ of the order of a few meters. Images of a solar eclipse are much rarer, but they allow to boost the oscillation baseline by around 8 orders of magnitude. In the remainder of this section we will discuss in detail how to analyse these two types of data sets.

\subsection{Analyzing the darks}
\label{sec:darks}

We begin our analysis by collecting all full-resolution darks, i.e.\ $2048\times2048$ pixels on a $35\times 35\,$ arcmin$^2$ field of view (FOV).\footnote{In principle there are also darks with lower resolution, but the zero-point subtraction in the calibration routine is optimised for the $2048\times2048$ images~\cite{2014SoPh..289.2781K}.} We then calibrate this data set using the ``empiric'' version of the XRT standard routine \cite{2014SoPh..289.2781K} of SolarSoft \cite{Freeland98} available at \url{https://www.lmsal.com/solarsoft/}. We find that the calibration to the zero point is quite precise with an uncertainty of about 1 data number (DN), which translates to a deposited energy of $208.05\,\mathrm{eV}$. For comparison, the uncalibrated images are scattered around 42\,DN. Finally, we apply a set of quality requirements to remove damaged, incomplete or unsuitable images (see appendix~\ref{sec:app-b} for details) which identify 960 valid images with a combined exposure of $t_\text{exp} = 9463\,\mathrm{s}$.

In agreement with Ref.~\cite{2014SoPh..289.2781K} we make the interesting observation that the average number of counts in a given dark is largely independent of the exposure, i.e.\ it does not depend on how long the detector was active. Likewise, there is no significant variation of the number of counts throughout the years of data taking. These findings suggest that external backgrounds like solar wind charge transfer~\cite{Wargelin:2004vm}, which should increase for longer exposure and vary with time, are subdominant and that the dominant contribution comes from internal sources. This contribution is expected to be handled well by the calibration routine.

Because Hinode XRT has to sustain the stress of constant exposure to the solar x-ray flux, its energy resolution cannot compete with more sensitive telescopes like XMM-Newton \cite{XMM:2001haf} or Chandra \cite{Weisskopf:2000tx}. In particular, Hinode's x-ray telescope was not designed to resolve individual photons but rather to measure a large flux of solar x-rays. Therefore, we only have access to the total energy deposited in each pixel over the exposure time of an image, which does not allow for directly distinguishing individual photon hits.\footnote{Note also the comment on photon counting in the data analysis guide \url{https://xrt.cfa.harvard.edu/resources/documents/XAG/XAG.pdf}.} Nevertheless, it is possible to restore some sensitivity to the spectral distribution if the exposure time of each image is small enough that one expects at most one photon hit per pixel under the signal hypothesis. We have verified a posteriori that this assumption is consistent with the sensitivity that we achieve. 

\begin{figure}[t]
    \centering
    \includegraphics[width=0.7\columnwidth]{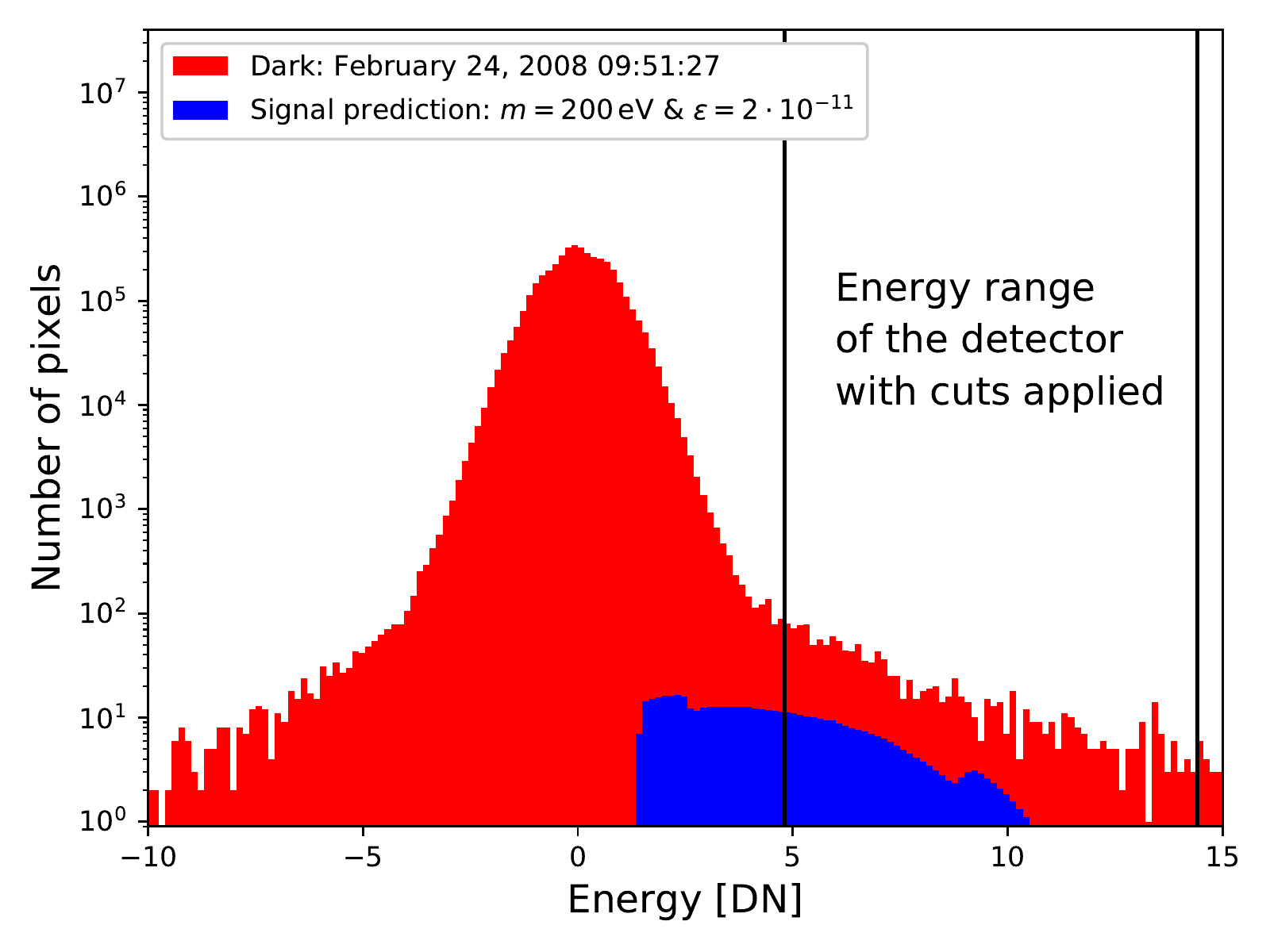}
    \caption{Histogram of a dark (red) which demonstrates the noise domination. The blue histogram shows the expected signal for a DP with $m=200\,$eV and $\epsilon=2\cdot 10^{-11}$. Cuts on the energy range to suppress noise are indicated by the black lines.}
    \label{fig:cuts}
\end{figure}

Using the calibration routine and its zero-point adjustment we can thus pick out potential hits that lie within the correct energy range.\footnote{See Ref.~\cite{Hudson:2012ee} for a different approach to extracting spectral information from the data.} 
The situation is illustrated in figure~\ref{fig:cuts}, which shows the distribution of energy deposition across all the pixels of a single dark. Thanks to the calibration, the distribution has a relatively narrow peak at a data number of zero, meaning that physical detector hits, i.e.\ external backgrounds and DP events, will show up as an excess in the region marked by the two black lines. The upper bound of this region ($E_\text{max} = 3\,\mathrm{keV} \approx 15\,\mathrm{DN}$) is given by the sensitivity limit of Hinode XRT, whereas the lower bound ($E_\text{min} =  1\,\mathrm{keV} \approx 5\,\mathrm{DN}$) is chosen well above the nominal threshold of $200\,\mathrm{eV}$ in order to suppress background while maintaining good sensitivity to a potential DP signal.

We also show in figure~\ref{fig:cuts} the signal prediction of a $200\,$eV DP for $\epsilon = 2 \cdot 10^{-11}$. We find that the signal is below the background level, meaning that it will not be possible to exclude the corresponding parameter point unless additional information on the spectral and angular distribution of both signal and background are included in the analysis. In the following we will discuss how this information can be included and show that this will enable us to significantly improve our sensitivity to DP signals.

We will refer to each pixel with a data number in the correct range as an ``event''. After stacking all darks, we find that for the vast majority of pixels the number of events per pixel follows a Poisson distribution with expectation value $\mu = 0.17$. A small fraction of around 0.1\% of the pixels, however, observes a much larger number of events, which can reach several hundred. These pixels are likely not functional and will be removed from the subsequent analysis (see appendix~\ref{sec:app-b} for details).

As a first step, we can now sum up the observed events in all pixels that point at the Sun (indicated by the white circle in the left panel of figure~\ref{fig:eclipses}). This number, called $S$, should be compared to the total signal prediction for the number of events
\begin{equation}
    R = t_\text{exp} \int_{0}^{R_\odot}\mathrm{d}r \int_{E_\text{min}}^{E_\text{max}}\mathrm{d}\omega \frac{\mathrm{d}\Phi}{\mathrm{d}\omega}\frac{\mathrm{d}X}{\mathrm{d}r}P(\text{DP}\rightarrow\gamma)Q(\omega)\; ,
\end{equation}
where $\mathrm{d}X/\mathrm{d}r$ can be obtained from eq.~\eqref{eq:fact} using $\mathrm{d}\Omega\approx 2\pi \psi \mathrm{d}\psi$ and $\psi\approx r / R_e$. $Q(\omega)$ denotes the combined efficiency of the detector, i.e.\ the effective area, the quantum efficiency, and the transmittivity of the filter \cite{2007SoPh..243...63G}. This prediction depends on the DP mass $m$ and the mixing parameter $\epsilon$.

\begin{figure}[t]
    \centering
    \includegraphics[width=0.9\columnwidth,trim={2cm 0 2cm 0}]{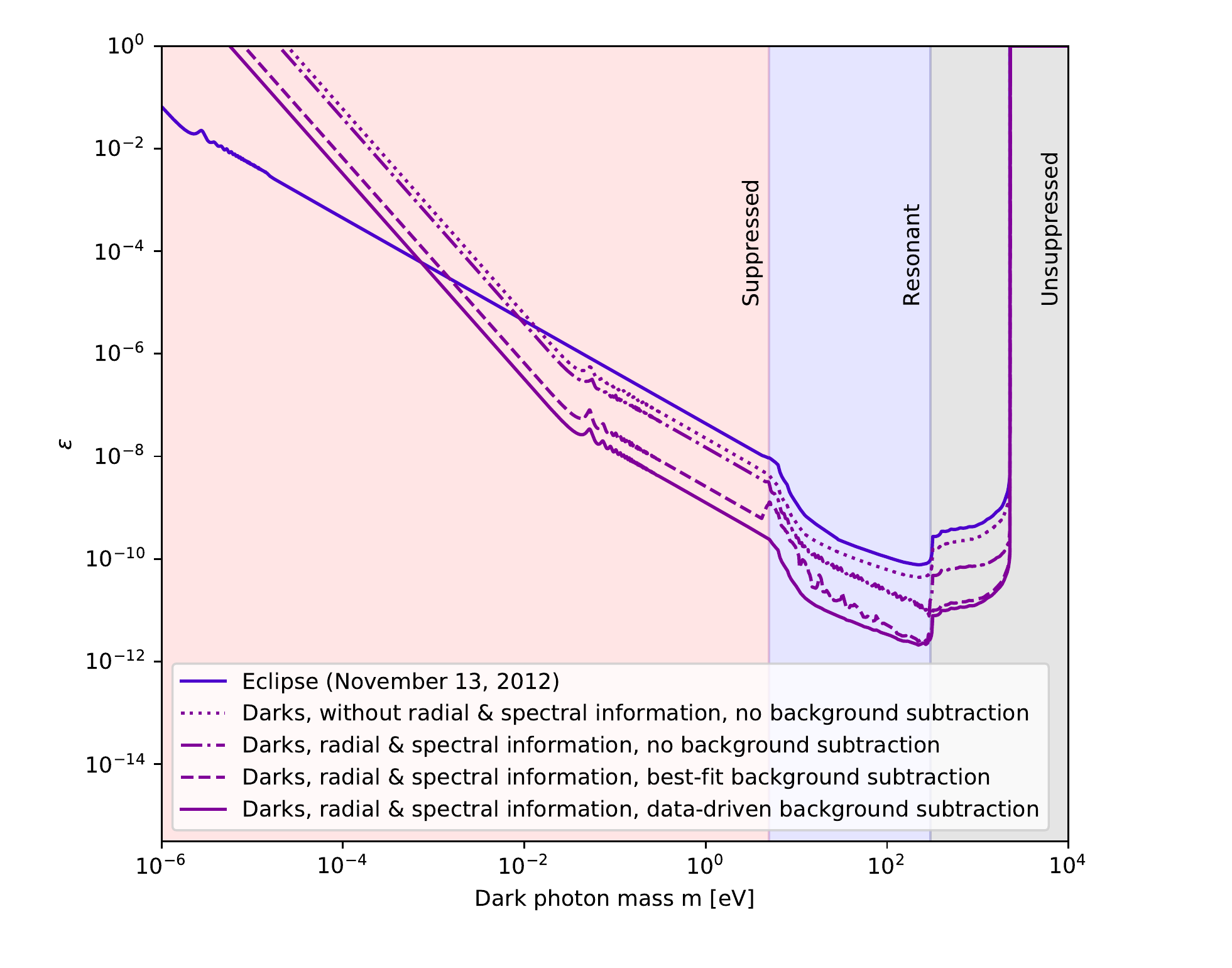}
    \caption{Constraints in the $m$--$\epsilon$ plane. In this plot, we focus on the different constraints derived in the previous paragraphs with emphasis on the improvements between naive and full analysis.}
    \label{fig:constraints}
\end{figure}

Based on the discussion above, we expect the total number of observed events to follow a Poisson distribution. Without any further information, we can therefore immediately exclude any signal hypothesis at 90\% confidence level, for which $R - 1.28 \sqrt{R} > S$, which (given that $R, S \gg 1$) is approximately equivalent to $R > S + 1.28 \sqrt{S}$. The exclusion limit obtained in this way is indicated by the dotted purple line in figure~\ref{fig:constraints}. We will refer to this method as the SHIPS/CAST-like analysis as it does not include any radial or spectral information of the signal prediction.

We now divide the signal region into 50 annuli $[r_j, r_{j+1}]$ of equal width, which enables us to perform a radial binning of the events. Each radial bin will be further divided into 10 energy bins $[E_i, E_{i+1}]$ of equal size. The observed total counts will be denoted by $S(E_i,r_j)$, while the signal prediction is given by
\begin{equation}
    R(E_i,r_j)=t_\text{exp}\int_{r_j}^{r_{j+1}}\mathrm{d}r \int_{E_i}^{E_{i+1}}\mathrm{d}\omega \, \frac{\mathrm{d}\Phi}{\mathrm{d}\omega}\frac{\mathrm{d}X}{\mathrm{d}r}P(\text{DP}\rightarrow\gamma)Q(\omega)\; .\label{eq:pred}
\end{equation}
This binning enables us to exclude any signal hypothesis for which the predicted number of events in any bin significantly exceeds observations. Assuming $S(E_i, r_j) \gg 1$ for all bins, we can define the test statistic 
\begin{equation}
    \chi^2(m, \epsilon) =\sum_{i,j}\frac{\text{max}\left(R(E_i,r_j)-S(E_i,r_j),0\right)^2}{S(E_i, r_j)} \; , \label{eq:chisq}
\end{equation}
where we take the maximum to ensure that only bins with $R(E_i, r_j) > S(E_i, r_j)$ contribute to the sum. We make this choice as the ``background-only'' hypothesis $\chi^2(\epsilon=0)$ would give a very large $\chi^2$ otherwise. We can then calculate an upper bound on the mixing parameter by solving
\begin{equation}
    \Delta \chi^2\equiv \chi^2(\epsilon)-\chi^2(\epsilon=0)=1.64\; ,\label{eq:conf}
\end{equation}
at $90\%$ confidence level.

\begin{figure}[t]
    \centering
    \includegraphics[width=1\textwidth,trim={2cm 0 2cm 0}]{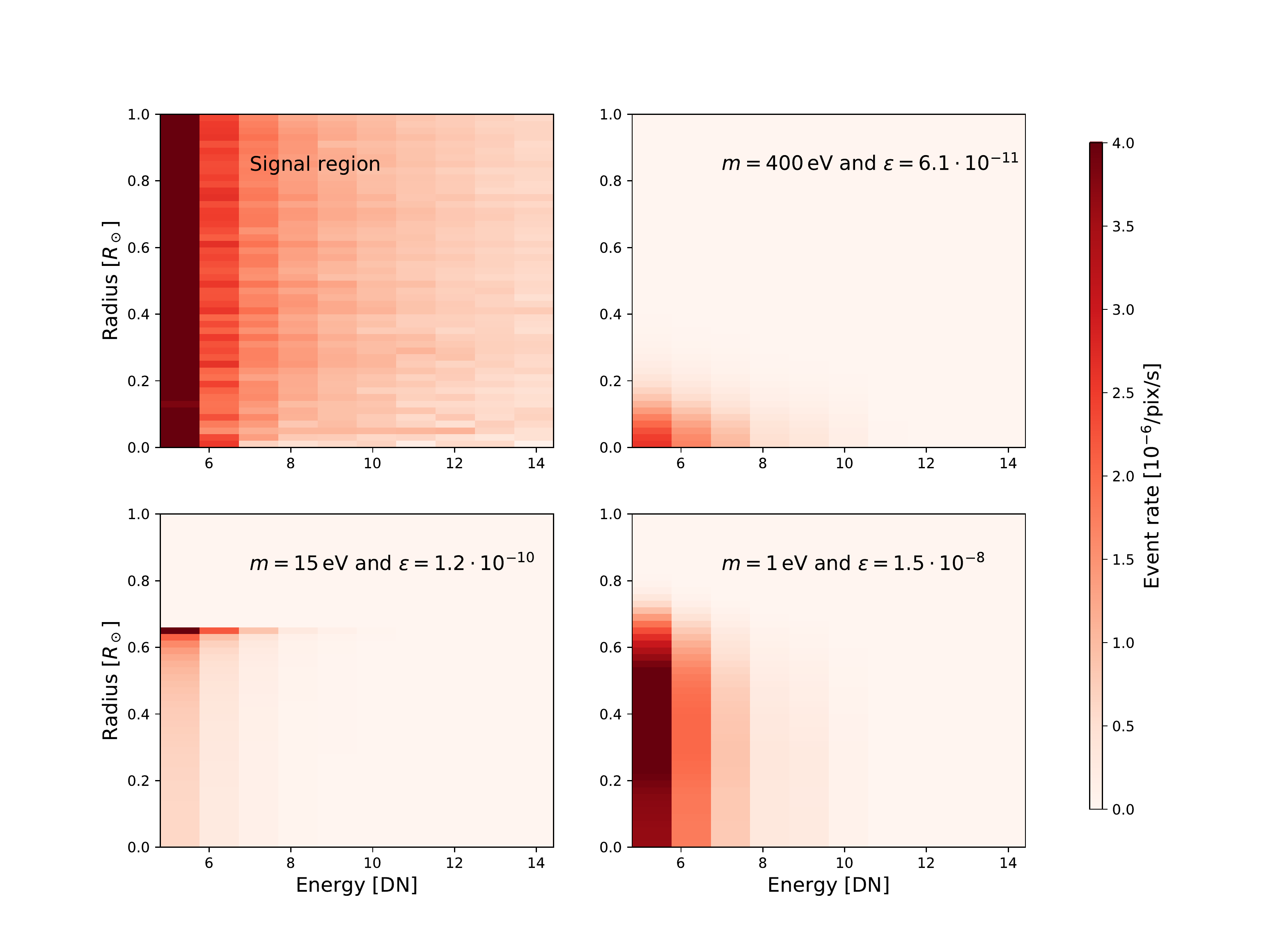}
    \caption{2D plot of the data in the signal region (top left) and the signal prediction for $400\,$eV (top right), $15\,$eV (bottom left), and $1\,$eV (bottom right) DP masses.}
    \label{fig:2dsignal}
\end{figure}

The exclusion limit obtained in this way is indicated by the dash-dotted purple line in figure~\ref{fig:constraints} and is found to be substantially stronger than the one obtained without angular and radial information. The reason is of course that the shape of signal and background distribution are very different, as demonstrated in figure~\ref{fig:2dsignal}. In the top-left panel we show the rather simple distribution of the data (in terms of events per pixel per second) across the signal bins. We see a clear decrease of the event rate towards larger energies, while the rate is relatively constant in the radial direction. 
 
For comparison we show signal predictions for three different DP masses ($1\,$eV (suppressed), $15\,$eV (resonant), $400\,$eV (unsuppressed)) in the remaining panels of figure~\ref{fig:2dsignal}. The values of the kinetic mixing parameter $\epsilon$ are chosen to lie close to the exclusion limit obtained above.
Clearly, in each of these cases, the total predicted number of signal events is significantly smaller than the total number of observed events, but an exclusion is possible using the differential information.

Another key observation from figure~\ref{fig:2dsignal} is that the background is flat up to large radii, whereas the signal peaks at smaller radii. This feature makes it possible to perform background subtraction in order to obtain even stronger bounds. In the following, we will discuss two possible approaches for background subtraction: using a fitting approach to the data in the signal region and a data-driven approach using the background counts in a control region.

Indeed, we can see in figure~\ref{fig:eclipses} that there are a large number of pixels that do not point at the Sun. Since no DP signal is expected in these pixels, we can use them as a control region (CR) to determine the background rate. Denoting the number of events in the control region in a given energy bin by $B(E_i)$, we can obtain a background prediction for a given signal bin by rescaling with the number of pixels:
\begin{equation}
 B_S(E_i, r_j) = B(E_i) \frac{N_\text{pix}(E_i, r_j)}{N_\text{pix,CR}} \; .
\end{equation}
Since the number of pixels in the control region is large compared to any signal region, we can neglect the systematic uncertainty of the background prediction resulting from Poisson fluctuations in the control region in the following.

\begin{figure}[t]
    \centering
    \includegraphics[width=0.8\textwidth]{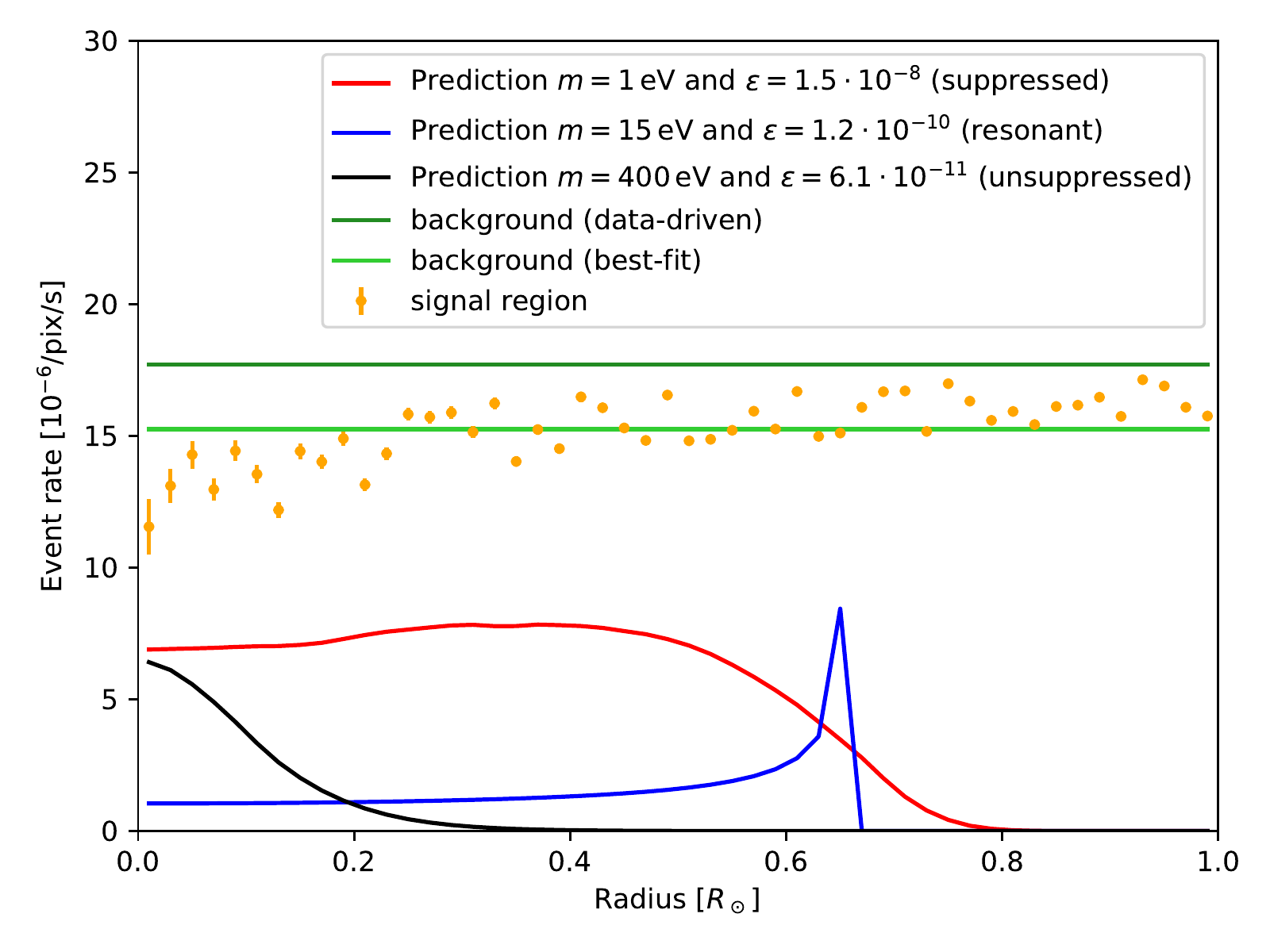}
    \caption{Data from background and signal region as well as the signal prediction for three masses. The values of $\epsilon$ have been adjusted to roughly match the signal region data before background subtraction and thus, they do not have any physical significance.}
    \label{fig:1dcomp}
\end{figure}

In figure~\ref{fig:1dcomp} we compare the background prediction obtained in this way to the actual observations as a function of radius (i.e.\ summed over all energy bins) and the corresponding signal predictions for the three benchmark points considered above. Again, we observe that the radial dependence of the signal is very different from the predicted background and the observed signal. The resonant region leads to a very distinct peak at the resonance very similar to figure~\ref{fig:shapes} while suppressed and unsuppressed regions again have a smoother radial dependence.\footnote{Note that the signal predictions look somewhat different from figure~\ref{fig:shapes} because they are now given per pixel and not per radial bin. Furthermore, the non-resonant predictions are estimated via eq.~\eqref{eq:weight} which means the angular distribution is weighted with the sensitivity of Hinode XRT and thus is instrument-dependent.} The vacuum-like case predicts a strong signal at small radii due to the high temperatures and densities, while for the suppressed scenario the prediction is more stretched out.

Figure~\ref{fig:1dcomp} also shows that the background prediction obtained from the control region is not perfect, in the sense that there is a slight decrease in the observed event rate towards the centre, such that the inferred background rate overpredicts the observation. To ensure that this effect does not introduce a bias in our analysis, we implement an alternative way to estimate the background, in which we simply fit the observed rate by a constant for each energy bin and multiply with $t_\text{exp}$ and $N_\text{pix}(E_i, r_j)$ to obtain the predicted number of background events called $B_S^\text{fit}(E_i, r_j)$.\footnote{In principle, one could also fit more complicated functions that capture the slight radial dependence of the observed rate. We have checked that doing so has no significant effect on our results.} Again, the systematic uncertainty of the background prediction is negligible compared to the statistical uncertainty in each signal bin.

Given a background estimate for each signal bin, we can define the test statistic
\begin{equation}
    \chi^2(m, \epsilon) =\sum_{i,j}\frac{\left(R(E_i,r_j)+B_S^\text{(fit)}(E_i,r_j)-S(E_i,r_j)\right)^2}{S(E_i,r_j)} \; . \label{eq:chisq2}
\end{equation} 
Since we now have a background model, this test statistic can in principle be used to search for a preference over the background-only hypothesis ($\epsilon = 0$). However, here we will limit ourselves to setting exclusion limits according to eq.~\eqref{eq:conf}. Since the signal prediction simply scales proportional to $\epsilon^4$, we can solve eq.~\eqref{eq:conf} analytically for fixed DP mass.

The exclusion limits obtained when using the data-driven (best-fit) background are given by the solid (dashed) purple line in figure~\ref{fig:constraints}. As expected, performing a background subtraction leads to bounds that are stronger by more than an order of magnitude in terms of $\epsilon$ (i.e.\ by more than four orders of magnitude in terms of the signal strength) compared to the naive approach. The two different methods to estimate the background give very similar results, confirming that our approach is robust.\footnote{For the data-driven background estimate, the background prediction is larger than the observed number of events in all bins, such that there is never a preference for a DP signal. For the best-fit background prediction, on the other hand, there are a few small excesses (see figure \ref{fig:1dcomp}), which can potentially be fitted by DPs in the resonant region, leading to slightly weaker exclusion limits.}   We conclude that using angular and spectral information to distinguish signal from background is essential to maximize the sensitivity of Hinode XRT. 

\subsection{Analyzing the eclipse}
\label{sec:eclipse}

Having analyzed the darks, let us now exchange large exposure for long oscillation baseline by considering solar eclipses in order to become more sensitive to smaller masses. In principle one could try to increase the exposure by stacking images of eclipses, but the rarity of these events does not leave us with many choices. After going through all eclipses in Hinode's lifetime, we identify the eclipses on March 19, 2007 and November 13, 2012 as the best candidates. We will focus on the latter as the corresponding images have better resolution ($1024\times 1024$ vs. $512\times 512$). Given the very similar solid angle covered by the Moon and Sun, there are very few images where the Sun is fully covered.  The most central of these images is shown in figure~\ref{fig:eclipses} (right).\footnote{There also exists a movie of this eclipse: \url{https://www.youtube.com/watch?v=XNzleYZL6u8}.}

It is clear from figure~\ref{fig:eclipses} (right) that both the control region and parts of the signal region are ``polluted'' by solar activity. Clearly, we can no longer rely on the control region for background subtraction. Looking at both raw and calibrated data, we observe a generic increase of events towards the outer regions of the image. This effect has also been observed in Ref.~\cite{Afshari_2016}, where it is used for the calibration of the telescope's point spread function. As a result it becomes necessary to remove the outermost parts of the signal region in our analysis.

In appendix \ref{sec:app-c}, we discuss how the eclipse image can be prepared for a statistical analysis analogous to the one applied to the darks. 
In principle, we could now perform the same binning in radius and energy as above. However, because of the short exposure, doing so would result in many empty bins, indicating that such a binning is unnecessarily fine. We therefore simply consider a single bin up to half of the solar radius to minimise the pollution from the solar corona. While we expect this selection to have very good signal efficiency in the suppressed and unsuppressed regime, we lose some sensitivity in the parameter region where the resonant peak occurs at large radius. We have checked that this decrease in sensitivity only mildly weakens our limits. 
The signal region contains 389 events, many of which can potentially be attributed to physical backgrounds or internal noise. Here we will not attempt to construct a background model and instead simply perform a SHIPS/CAST-like analysis without background subtraction as discussed above. Despite this simplification it is crucial to know the radial distribution of the DP signal in order to calculate the fraction of events that end up in the signal region. 

The resulting limits are shown in blue in figure~\ref{fig:constraints}. As expected, the exclusion limits obtained from the eclipse are subdominant for heavy DP masses, where the oscillation length is tiny compared to the size of the telescope. Nevertheless, the limit from the eclipse data is comparable to the one obtained from the darks without angular and spectral information and without background subtraction. This finding illustrates that the additional refinements discussed in section~\ref{sec:darks} are necessary in order to benefit from the large exposure of the darks.
On the other hand, for DP masses below around $1\,\mathrm{meV}$, the limits from the eclipse become stronger than the ones obtained from the darks, because the latter are suppressed as soon as the oscillation length becomes comparable to the size of the telescope. 

\begin{figure}[t]
    \centering
    \includegraphics[trim={2cm 0 2cm 0},width=0.9\textwidth]{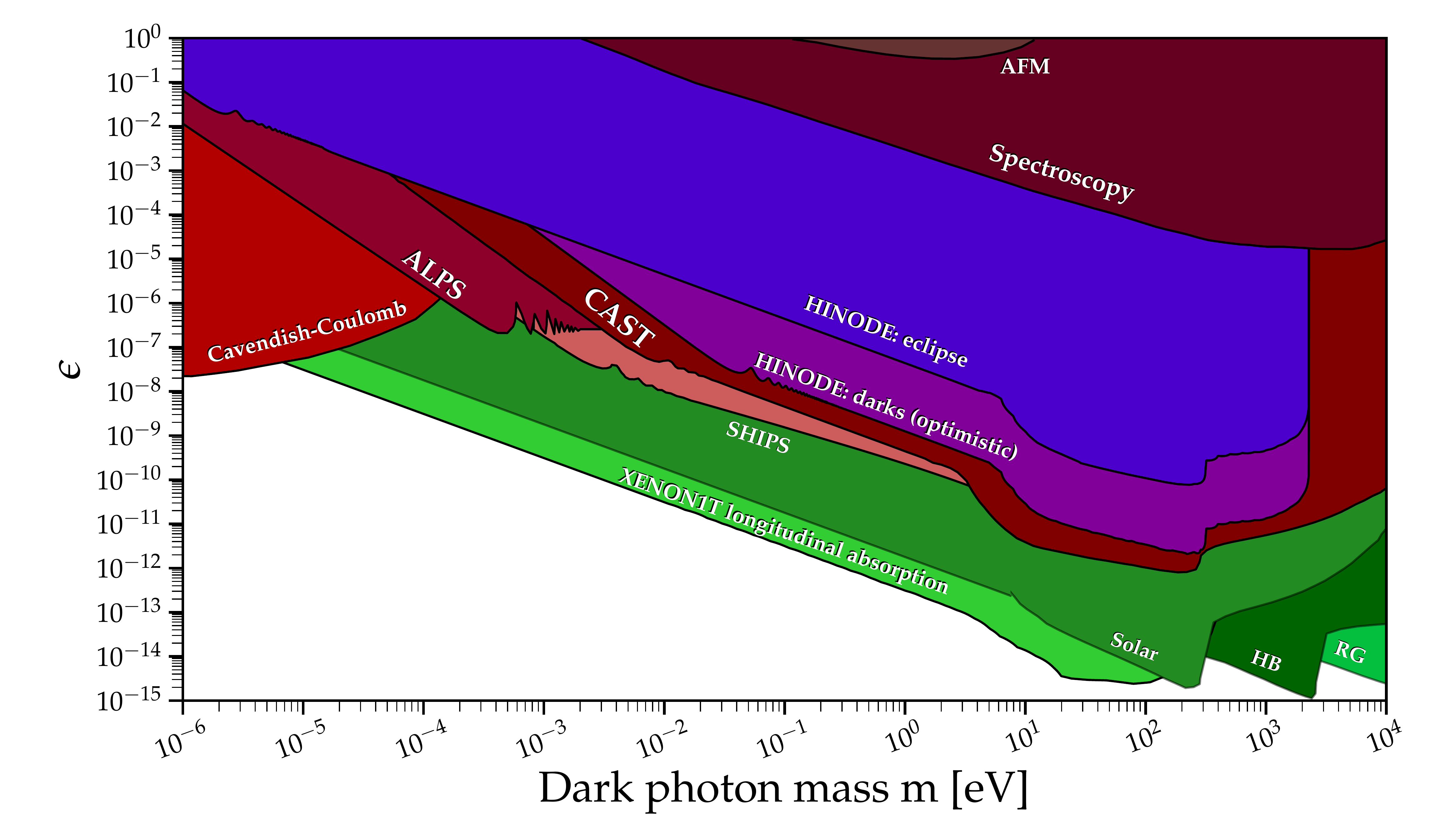}
    \caption{Same as figure~\ref{fig:constraints}, but comparing the limits obtained in this work to existing constraints on DPs from the literature (see Ref~\cite{Caputo:2021eaa}). Constraints coloured in green rely mostly or even completely on the production of longitudinal DPs in the Sun, whereas those coloured in red are derived from various other DP effects. For the analysis of the darks, we only show our most optimistic exclusion limit, obtained using data-driven background estimates.}
    \label{fig:constraints_ohare}
\end{figure}

\subsection{Discussion}
\label{sec:disc}

In figure~\ref{fig:constraints_ohare}, we put our constraints into a broader context. The two limits labeled Hinode XRT correspond to the most optimistic analysis of the darks (i.e.\ using data-driven background subtraction) and the analysis of the eclipse data and correspond to the solid purple and blue line in figure~\ref{fig:constraints}. We use the excellent database and layout from Ref.~\cite{Caputo:2021eaa} (available through Ref.~\cite{AxionLimits}) to compare our results to a wide range of existing constraints. As mentioned in section~\ref{sec:oscillation}, the best limit in our region of interest is given by the longitudinal production, here shown in terms of the solar cooling argument (``solar'') and the searches for absorption in DM direct detection experiments discussed above (``XENON1T longitudinal absorption'') \cite{XENON:2021nad} which we have added by hand. Nevertheless, we conclude that Hinode XRT clearly outperforms lab experiments like spectroscopy searches \cite{Jaeckel:2010xx}.\footnote{Note that this work also includes a very nice discussion of the subtle reason why many lab searches become insensitive to the DP for smaller masses.}

Further, we observe that LSW experiments, here represented by ALPS \cite{Ehret:2010mh}, become important for small DP masses because they do not need to track the Sun and therefore can construct larger instruments. The next generation of experiments, including ALPS II~\cite{Spector:2019ooq}, will be able to explore even more parameter space. An additional advantage of these experiments is the control over the initial state, thus eliminating the astrophysical uncertainties that helioscope searches potentially suffer from.

\begin{table*}[t]
\center
 \begin{tabular}{ p{5cm} p{2cm} p{2cm} p{2cm} p{2cm} } 
 \hline
 \hline
Experiment & CAST &  SHIPS &  Darks & Eclipse  \\
\hline
Typical energy in eV & 5000 & 3 & 1000 & 1000  \\
Length in m & 9.26 & 5 & $\sim$ 1 & $\sim 3.8\cdot 10^8$\\
Kink mass in eV & $7\cdot 10^{-3}$ & $2\cdot 10^{-3}$  & $3\cdot 10^{-2}$  & $2\cdot 10^{-6}$  \\
Exposure times in s & 709200 & 1188000 & 9500 & 1 \\
Sensitivity to $\epsilon$ at $m=1\,$eV & $6\cdot 10^{-10}$ & $3\cdot 10^{-10}$ & $1\cdot 10^{-9}$ & $4\cdot 10^{-8}$\\
\hline
 \hline
\end{tabular}
\label{tab:osc}
\caption{Specifications of different solar DP searches.}
\end{table*}

In comparison with dedicated helioscope experiments, namely CAST and SHIPS, the constraints obtained from Hinode XRT are slightly weaker. This difference stems from a combination of the precise instruments and the unmatched exposure used by CAST and SHIPS (see table~\ref{tab:osc}). Nevertheless, the fact that Hinode XRT can be competitive with much smaller exposure shows that additional sensitivity to angular and spectral information is a similarly important asset for a helioscope. 

The largest DP mass accessible by the different helioscopes is set by the maximum energy that can be probed by each instrument. As CAST is sensitive up to $15\,$keV, it wins in this comparison, especially compared to SHIPS, which operated in the visible region and therefore loses sensitivity already at eV scales. We also note the clear bump structure observed for the two x-ray experiments for DP masses in the range $\sim 5$--$300$~eV, corresponding to resonant DP production in the Sun. The loss of sensitivity at small DP masses, on the other hand, depends on the typical oscillation length. The kink in the exclusion limit corresponds to the point when the argument of the sine in eq.~\eqref{eq:oscvac} becomes $\mathcal{O}(1)$:
\begin{equation}
    \frac{m^2L}{4\omega}\sim 1  \Leftrightarrow m\sim \sqrt{\frac{4\omega}{L}}.\label{eq:kink}
\end{equation}
The relevant scales for each experiment can be found in table \ref{tab:osc}. Clearly, the eclipse data offers an unparalleled reach in terms of small DP masses.

\section{Conclusions}
\label{sec:conc}

The Sun as a laboratory for light and weakly coupled BSM physics has been studied extensively. In the context of dark photons (DPs), there are barely any other constraints that can compete with the ones obtained from solar DP production in the eV to keV mass range. 
In this work we have studied how the sensitivity to such DPs can be maximised by exploiting the available information on the angular and spectral distribution of DPs produced in the Sun. For this purpose, we have developed a new way to estimate the angular distribution of DPs in the case that resonant production in the Sun is not possible. We have then used publicly available data from the solar x-ray telescope Hinode XRT with excellent angular resolution to reveal the potential to improve constraints on DPs.

First we have considered so-called darks, i.e.\ calibration images with closed shutter, which effectively make the telescope a space-based helioscope. Given the very low count rates, we were able to generate spectral information from these images by using the standard calibration procedure. We then developed and applied selection cuts to remove damaged images and pixels with large intrinsic background and identified two possible procedures for background subtraction. We find that the combination of including angular and spectral information and subtracting backgrounds leads to an improvement upon naive constraints from ``event counting'' by more than an order of magnitude in terms of the kinetic mixing parameter.

Furthermore, we have demonstrated that solar x-ray telescopes can also act as a giant helioscope when observing a solar eclipse. Doing so increases the sensitivity to much smaller DP masses at the expense of substantially reducing the exposure. Using a single image of a total eclipse, we have obtained an exclusion limit that outperforms existing helioscopes for DP masses below 1 meV, although it cannot compete with exclusions from LSW experiments like ALPS, because the production of very light DPs in the Sun is suppressed due to the large plasma frequency.

We emphasise that for this proof-of-principle analysis we have used very simple background models, which do not fit all trends in the data and could easily be refined, in particular for the case of the eclipse data. Improving the analysis routine further and including also partial eclipses should make it possible to increase the exposure by an order of magnitude or more. Increasing the exposure even further appears challenging due to the rareness of solar eclipses. 
An exciting possibility would be to use a telescope on a lunar-centric orbit, thereby sacrificing some of the oscillation baseline for much longer exposures.

The analysis routine that we have developed can also be used to study the angular distribution of solar DPs in the most recent version of the CAST experiment and in the future helioscopes BabyIAXO and IAXO~\cite{IAXO:2020wwp,Lakic:2020cin}. In this context, it would also be warranted to further improve the calculation of the DP angular distribution, including an update of the solar model~\cite{Turck-Chieze:2016vbl,Vinyoles:2016djt},  and estimating the corresponding uncertainties. 
Given the plans to measure the polarisation of solar x-rays with the CUSP project~\cite{2022arXiv220806211F}, it would also be highly interesting to obtain predictions for the polarisation of DPs produced in the Sun, which would offer a completely new lever to improve the limits (see also Ref.~\cite{Caputo:2021eaa}).
The combination of all of these efforts will enable us to probe new parameter space for the transverse components of dark photons in the near future. 

\acknowledgments{We would like to thank Thomas Reiprich for helpful discussions and Javier Redondo for providing the tabulated dark photon predictions. This work is funded by the Deutsche Forschungsgemeinschaft (DFG) through the Emmy Noether Grant No. KA 4662/1-1 and Germany’s Excellence
Strategy -- EXC 2121 ``Quantum Universe'' -- 390833306.
Hinode is a Japanese mission developed and launched by ISAS/JAXA, with NAOJ as domestic partner and NASA and STFC (UK) as international partners. It is operated by these agencies in co-operation with ESA and NSC (Norway).}

\appendix
\section{Details on the production of solar DPs}
\label{sec:app-a}

In this appendix we present some more technical details regarding the calculation of the solar DP flux.

In order to derive the full distribution in the resonant case we begin with the derivation of the spectral part, i.e.\ we have to explicitly perform the calculation in eq.~\eqref{eq:total} using the narrow resonance approximation:
\begin{equation}
    P(\omega,r,\theta)=\frac{\epsilon^2 m^4}{(m_\gamma^2(r)-m^2)^2+(\omega \Gamma(\omega,r))^2} \xrightarrow{\Delta r_\ast \ll R_\odot}\frac{\pi\epsilon^2m^4\delta(r-r_\ast)}{\omega \Gamma(\omega)\left|\frac{\mathrm{d}m_\gamma^2}{\mathrm{d}r}\right|} \; ,
\end{equation}
where $\Delta r_\ast = \omega\Gamma(\omega,r_\ast) \left|\mathrm{d}m_\gamma^2/\mathrm{d}r\right|_{r_\ast}^{-1}$ defines the sharpness of the peak.\footnote{We note that there is a hierarchy in length scales here: while the plasma changes slowly on scales of the oscillation length $l_\text{plasma}$, the region where the resonance condition is met is small in comparison to the full radial extent of the Sun, i.e.\ $l_\text{plasma}\ll\Delta r_\ast \ll R_\odot$.}
Using this approximation, we find
\begin{align}
\frac{\mathrm{d}\Phi}{\mathrm{d}\omega}    \approx& \frac{\omega\sqrt{\omega^2-m^2}}{2\pi^2 R_e^2}\int_0^{R_\odot} r^2\mathrm{d}r \int_{-1}^1 \mathrm{d}\cos\theta \frac{\Gamma (\omega,r)}{e^{\omega/T(r)}-1}\frac{\pi\epsilon^2m^4\delta(r-r_\ast)}{\omega \Gamma(\omega,r_\ast)\left|\frac{\mathrm{d}m_\gamma^2}{\mathrm{d}r}\right|_{r_\ast}}\nonumber\\
    =& \epsilon^2 m^4\frac{r_\ast^2}{R_e^2}\frac{\sqrt{\omega^2-m^2}}{\pi\left(e^{\omega/T(r_\ast)}-1\right)\left|\frac{\mathrm{d}m_\gamma^2}{\mathrm{d}r}\right|_{r_\ast}}\; .\label{eq:spectral}
\end{align}
This expression agrees with Ref.~\cite{2008} except for a factor of $\frac{\pi}{2}$ resulting from a mistake in the implementation of the narrow resonance approximation.

With this result at hand, we can now evaluate eq.~\eqref{eq:truangu}:
\begin{align}
       \frac{\mathrm{d}\Phi}{\mathrm{d}\omega \mathrm{d}\Omega}\approx&  \frac{\omega\sqrt{\omega^2-m^2}}{4\pi^3}\int_{r_\text{min}}^{R_\odot} \frac{2r \mathrm{d}r}{\sqrt{r^2-r_\text{min}^2}} \frac{\Gamma (\omega,r)}{e^{\omega/T}-1}\frac{\pi\epsilon^2m^4\delta(r-r_\ast)}{\omega \Gamma(\omega)\left|\frac{\mathrm{d}m_\gamma^2}{\mathrm{d}r}\right|}\nonumber\\
    =&\left(\epsilon^2 m^4 \frac{r_\ast^2}{R_e^2} \frac{\sqrt{\omega^2-m^2}}{\pi(e^{\omega/T}-1)\left|\frac{\mathrm{d}m_\gamma^2}{\mathrm{d}r}\right|}\right)\left(\frac{1}{2\pi}\frac{R_e}{r_\ast} \frac{1}{\sqrt{\psi_\ast^2-\psi^2}}\Theta(r_\ast-r_\text{min})\right)\; ,
    \end{align}
which leads to the factorisation given in eq.~\eqref{eq:factorisation}. To conclude this appendix, we confirm that our result is correctly normalised:
\begin{equation}
\int \mathrm{d}\Omega\frac{\mathrm{d}X}{\mathrm{d}\Omega}\approx\frac{2\pi}{2\pi}\int_0^{\psi_\ast}\frac{R_e}{r_\ast}\frac{\sin\psi \mathrm{d}\psi}{\sqrt{\psi_\ast^2-\psi^2}}=\frac{R_e}{r_\ast} \psi_\ast=1\; ,
\label{eq:norm}
\end{equation}
where we have used that $\psi\ll 1$ over the whole range of integration. In contrast, the corresponding expression from Ref.~\cite{2015} is not correctly normalised, although the shape of the distribution is similar to ours. 

\section{Discussion of image selection and choice of cuts}
\label{sec:app-b}

In this appendix, we will discuss the details of how to select events from the calibrated images.
\begin{figure}[b]
    \centering
    \includegraphics[width=0.8\columnwidth]{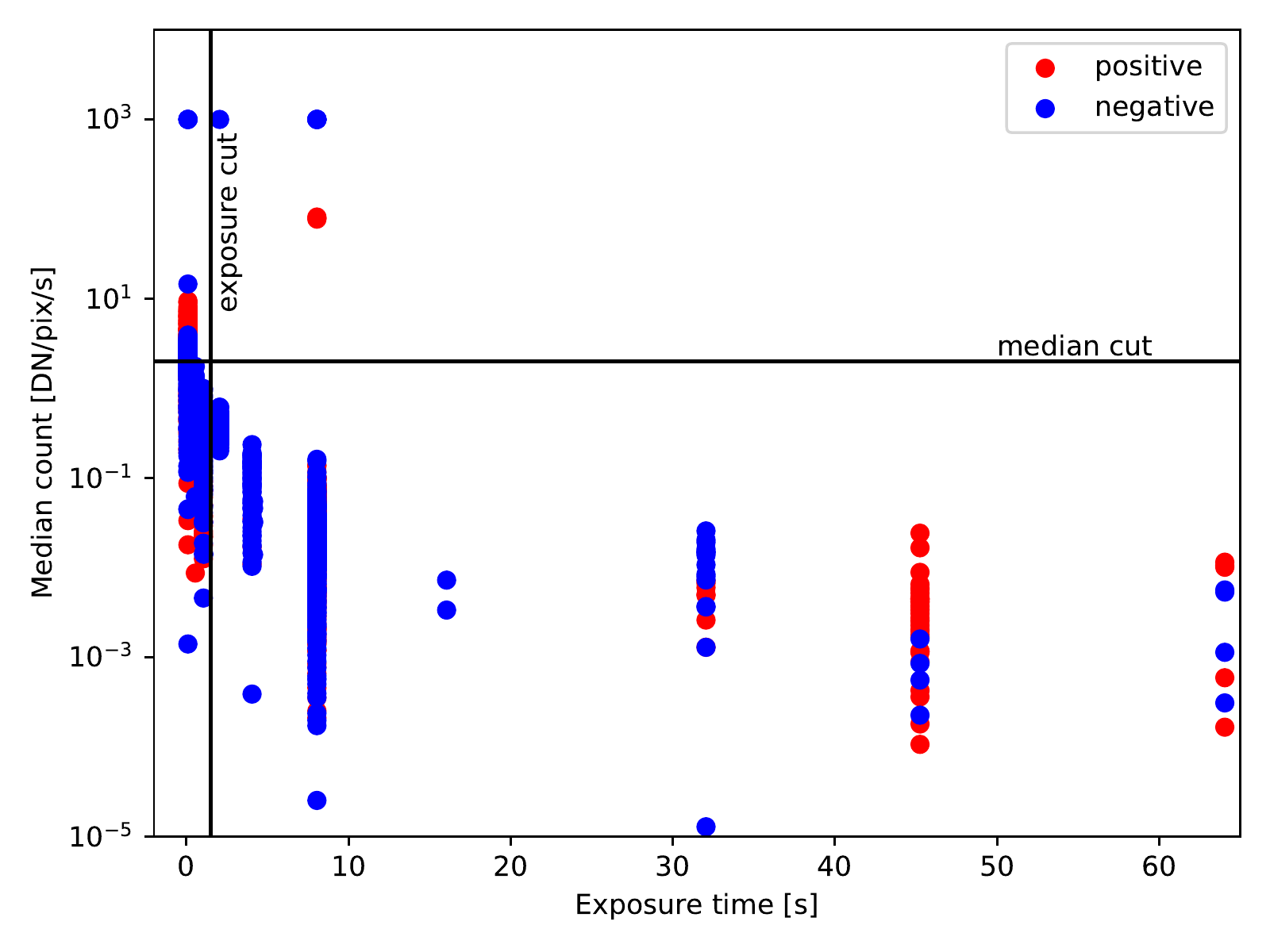}
    \caption{Scatter plot of all darks in the plane of exposure versus median count rate. We show the sign of the median in red/blue to allow for a logarithmic plot. High-quality images with successful calibration can be selected by applying the two cuts indicated by the horizontal and vertical line.}
    \label{fig:scatter}
\end{figure}
\begin{figure}[t]
    \centering
    \includegraphics[width=0.49\columnwidth]{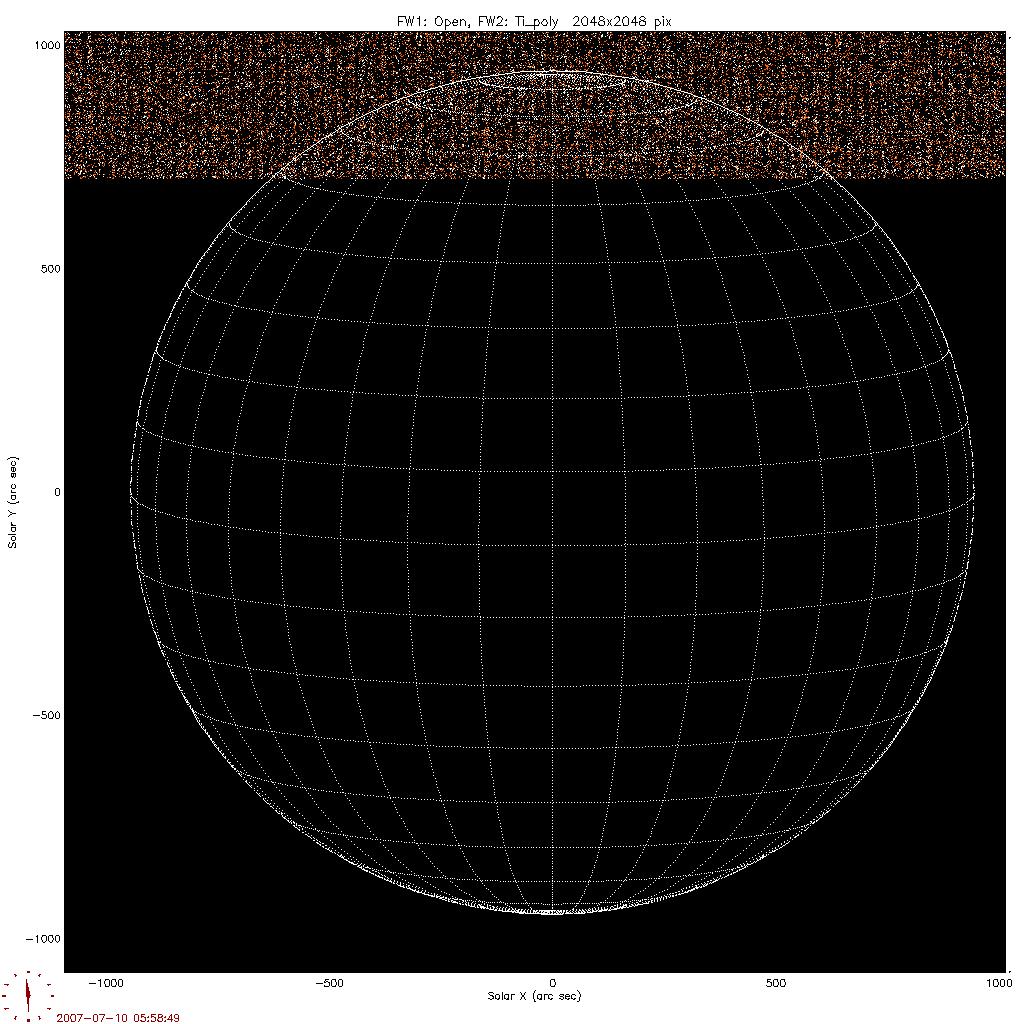}
    \includegraphics[width=0.49\columnwidth]{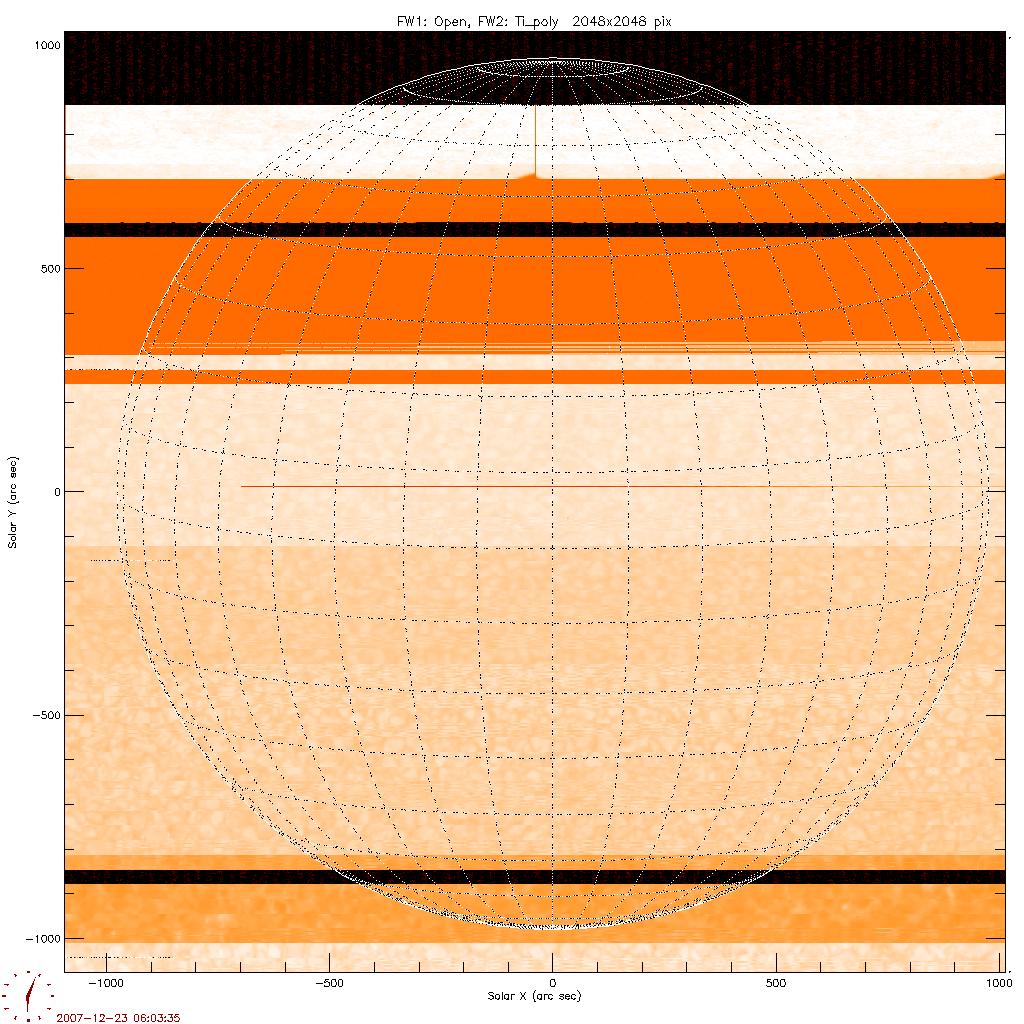}
    \caption{Images from \url{http://sdc.uio.no/search/form}. Left: Image with an enormous number of empty pixels. Right: Image that is obviously damaged but not necessarily empty. Both images are removed by our quality selection cuts.}
    \label{fig:damage}
\end{figure}

First of all, we need to identify potentially damaged images. As damaged or oversaturated pixels are given the value $-999$ in the calibration process, it is rather simple to discard strongly damaged images. We implement this by confirming that the absolute value of the median is smaller than $1\,$DN/pix/s which acts as a solid indicator for mostly intact images. 
Furthermore, we throw away images with exposure below $1.5\,$s as these tend to fluctuate more than images with long exposure. We use this combination of cuts because only for images with more than $1.5\,$s exposure the median is consistently and clearly below the median cutoff if we ignore obvious outliers, see figure~\ref{fig:scatter}. We also note that removing these images from the analysis only reduces the total exposure by a few percent. Figure \ref{fig:eclipses} shows a viable image (left) and figure \ref{fig:damage} (left) shows an image that is rejected due to missing pixels.

\begin{figure}[t]
    \centering
    \includegraphics[width=0.75\columnwidth]{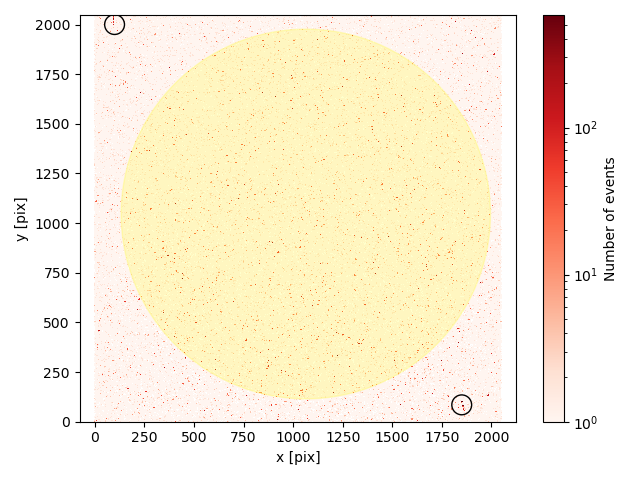}
    \caption{If we collect all viable events in a 2D histogram we can clearly observe two strong streaks (encircled) that will systematically lead to larger event rates in the background region (light red shading) than in the signal region (light yellow shading).}
    \label{fig:2dhist}
\end{figure}

\begin{figure}[t]
    \centering
    \includegraphics[width=0.75\columnwidth]{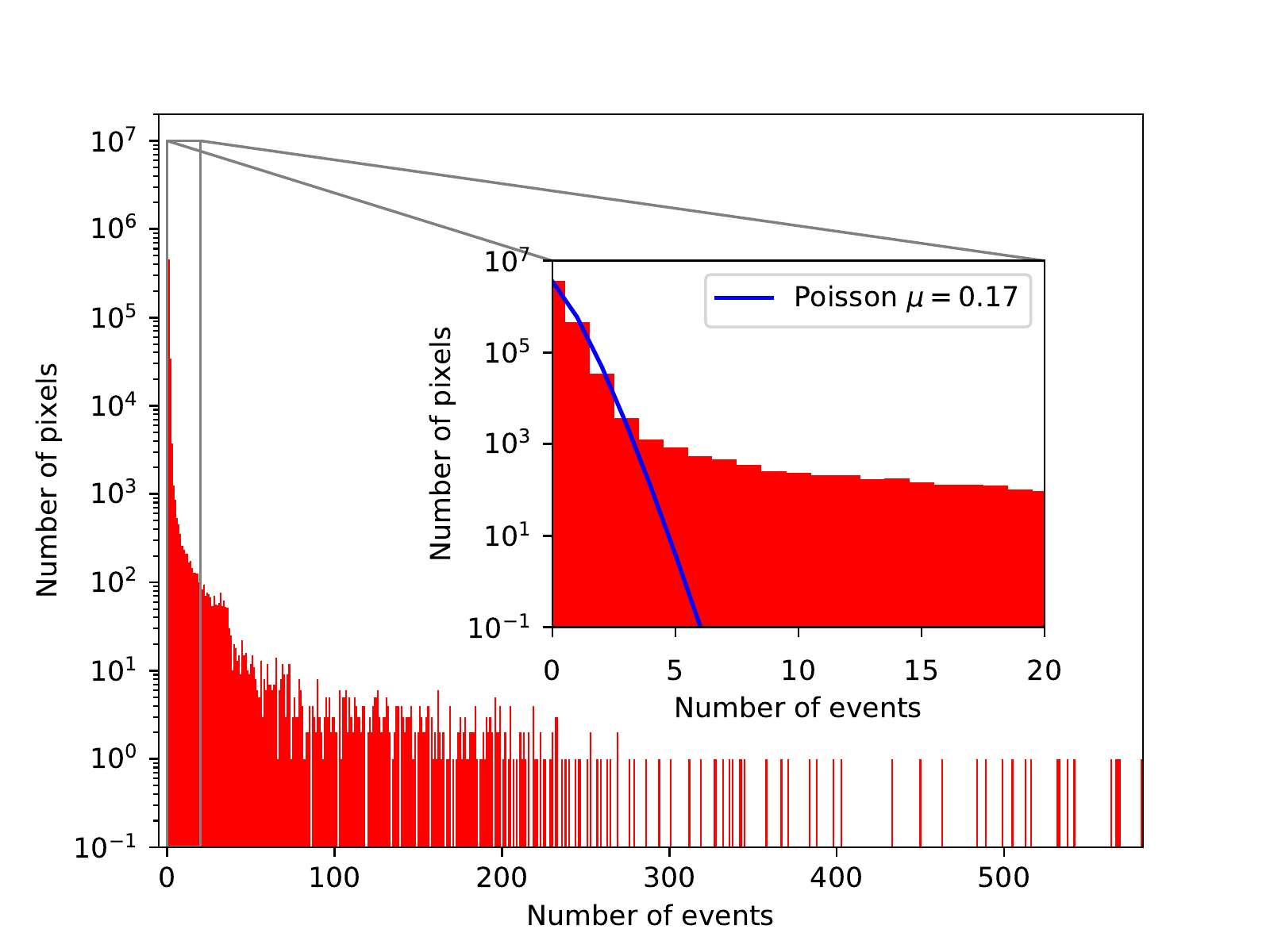}
    \caption{Histogram of the number of events observed in each pixel after stacking all darks. The distribution extends up to almost 600 events per pixel. This is to be contrasted with the (expected) Poisson distribution shown in the inset.}
    \label{fig:1dhist}
\end{figure}

We find that these cuts still let some obviously damaged images like the one in figure~\ref{fig:damage} (right) pass. These can be removed by constraining the maximum allowed number of events per image, which acts as a relatively secure method to identify strongly damaged images. We allow at most 9900 counts in the background bin and at most 990 in any of the radial signal bins.

Even with these extensive measures, we still observe some systematic trends in these images. For example, the number of events in the inner radial bins contain (on average) significantly fewer events per pixel than the outer and background bins. Part of this effect can be attributed to a small number of pixels with anomalously high number of counts. This is illustrated in the 2D histogram figure~\ref{fig:2dhist}, which shows the number of events in each pixel when summing up all 960 viable images. In the upper left and lower right corner, there are clearly visible streaks with several hundred of events in only a few pixels (compared to an average of 0.17 events per pixel). Several more such unexpected substructures can be identified across the image.

Fortunately, pixels with anomalously high count rates can be readily identified by looking at the 1D histogram of event counts in figure~\ref{fig:1dhist}. In addition to the Poisson distribution with $\mu = 0.17$ we observe a long-stretched tail of events up to a very large number of events. Given that the Poisson distribution predicts that there should be no pixels with more than five events, we remove any such pixels from the stacked image. This cut is quite efficient in that it removes only about $0.1\%$ of all pixels while reducing the total number of observed events by $23\%$.

Removing anomalous pixels reduces the systematic trend in the data but does not eliminate it completely. 
Even with the distribution now being close to Poissonian, we still observe on average a larger number of events in the outer bins. It is conceivable that this trend in fact reflects a physical effect that should be included in a more refined background model. A deeper analysis of this issue is beyond the scope of the present work.

\section{Calibration for the eclipse analysis}
\label{sec:app-c}

In this appendix, we discuss how to prepare the eclipse data such that it can be analysed using the same statistical methods as for the darks.
\begin{figure}[b]
    \centering
    \includegraphics[width=0.8\columnwidth]{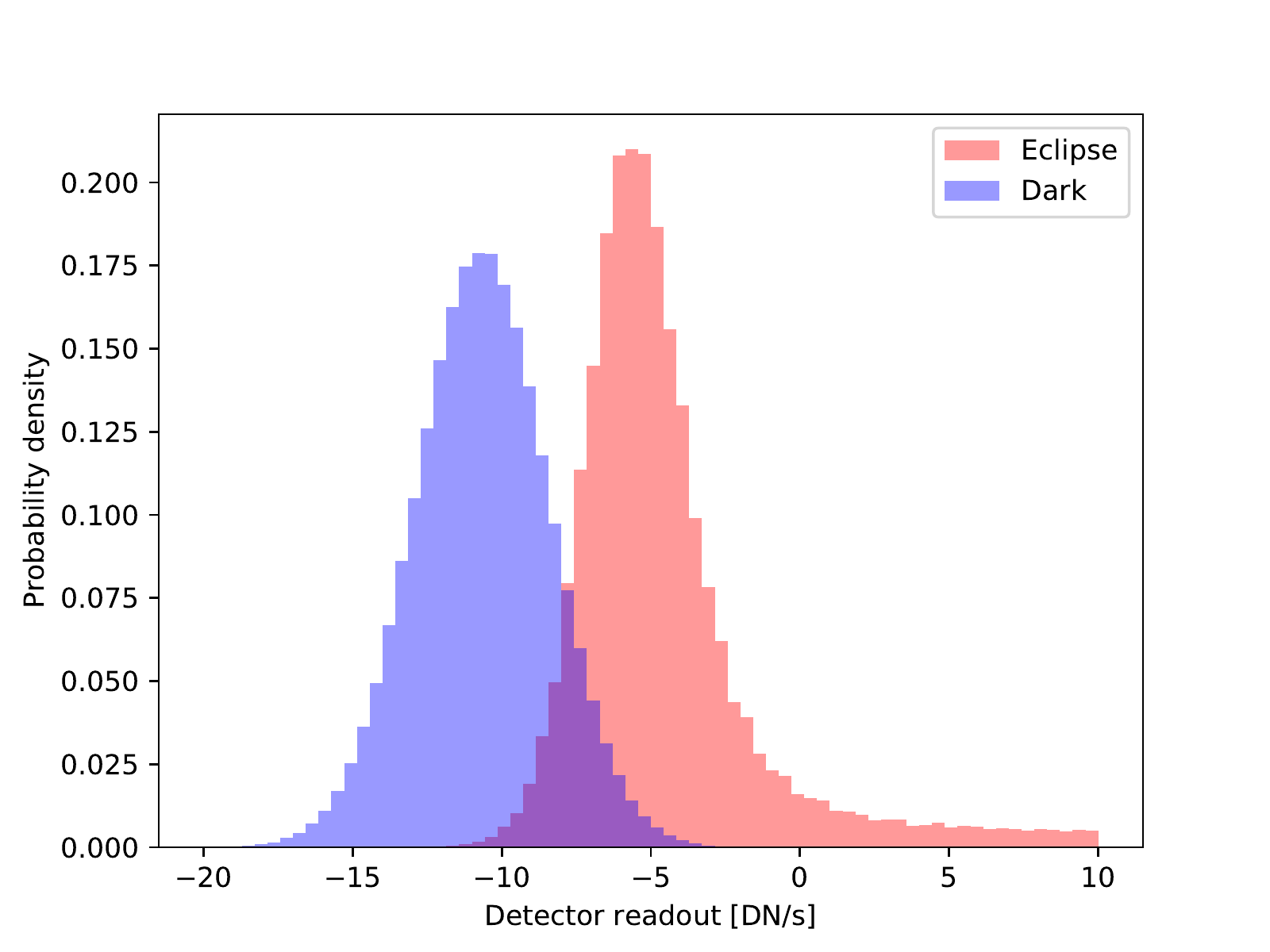}
    \caption{Normalized histogram of the DN/s for the eclipse (red) and a dark with the same resolution taken on the same day (blue). We observe that the zero-point calibration failed for both images and led to very different results. In the image of the eclipse one can clearly see the non-Gaussian tail from the solar x-rays on top of the Gaussian internal noise.}
    \label{fig:darkvseclipse}
\end{figure}
We find that for the image of the eclipse the calibration routine is not successful, such that we end up with a clear deviation from the expected scattering around zero (see figure~\ref{fig:darkvseclipse}). Unfortunately, it is also not possible to simply subtract a dark taken on the same day and with the same resolution, because the zero-point calibration yields very different results for the two images.\footnote{Moreover, a direct subtraction of a dark would also subtract any potential DP signal in the mass range where the closed telescope acts as a helioscope, i.e.\ when the oscillation length is short.}

Nevertheless, we can use the inner regions of the image to determine the physical zero point of the data number. To this end, we observe that the distribution of DNs for the dark (blue histogram in figure~\ref{fig:darkvseclipse}) looks Gaussian to good approximation, whereas the image of the eclipse exhibits a clear non-Gaussian tail of pixels with large energy depositions due to the solar x-rays from the regions of the Sun not covered by the Moon (red histogram). Nevertheless, in the central region of the image, where the solar x-rays are completely blocked by the Moon, the dominant contribution arises from internal noise leading to the approximately Gaussian peak.

\begin{figure}[t]
    \centering
    \includegraphics[width=0.7\columnwidth]{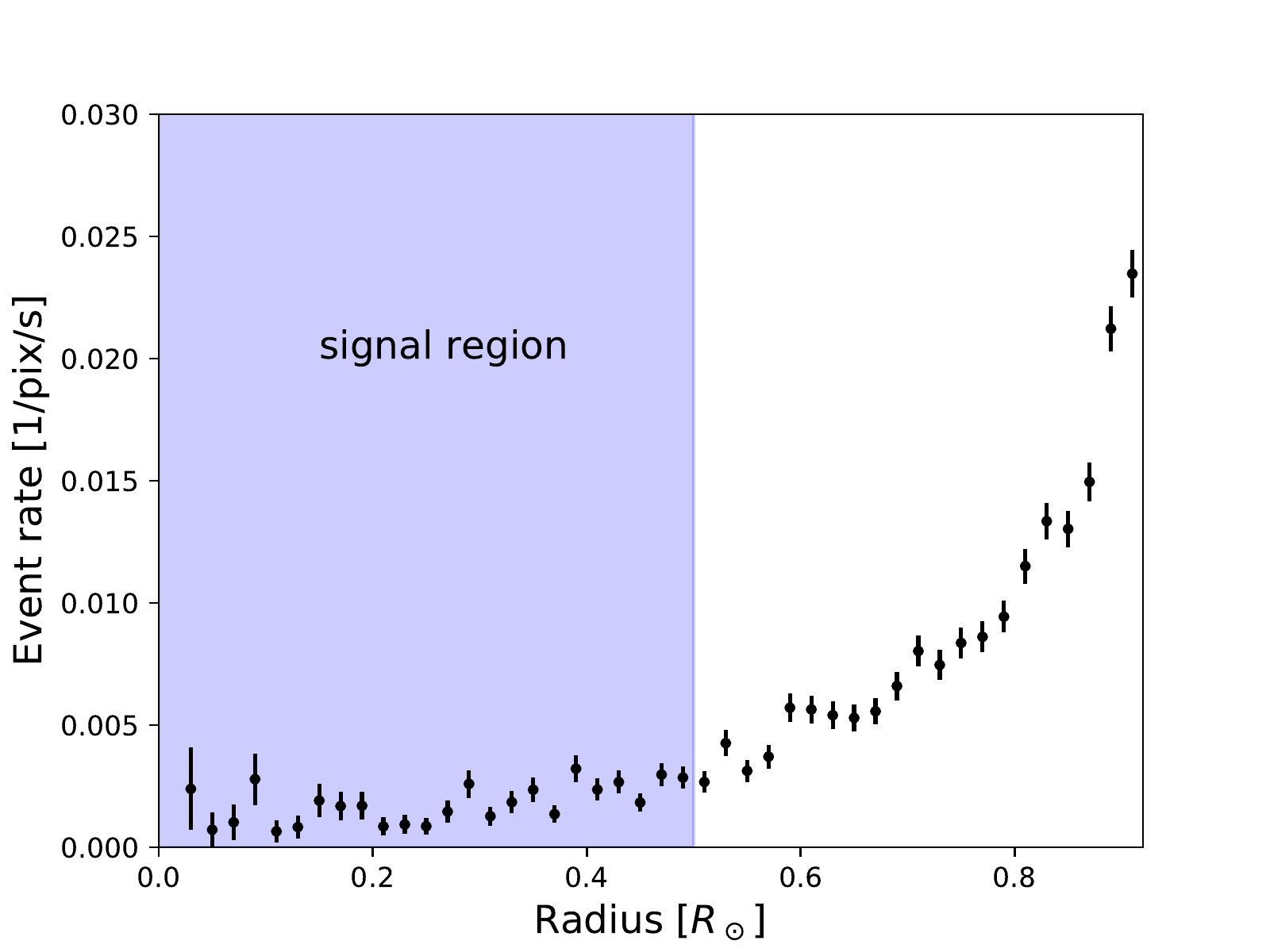}
    \caption{Plot of the event rate as a function of radius for the eclipse image. We only use the signal region for our zero-point determination.}
    \label{fig:zero}
\end{figure}

When considering only pixels within the inner 50\% of the solar radius, we find that the distribution is Gaussian to good approximation. We can thus use these pixels to determine the zero point to be at $-5.16$ DN/s. We show the radial dependence of the event rate after performing the manual zero-point calibration in figure~\ref{fig:zero}, which clearly shows the strong increase of the event rate towards large radii. Comparing this figure to figure~\ref{fig:1dcomp} one can also immediately see that the overall event rate (and the error bars) are orders of magnitude larger as a result of the contamination with solar x-rays and the limited statistics. 

After this dedicated calibration, we perform a consistency check on the effectivity of the cuts. Even if we identify pixels with a deposited energy larger than $3000\,$eV as hits by at least two background photons of any energy within the sensitivity range of the detector then the fraction of pixels with potential pollution stays below $5\%$ even in the most outer radial bin. In general, such a large energy deposition is very unlikely to originate from internal noise and is therefore interpreted as the physical background from the solar corona. We conclude that our cuts remove the problematic region and leave us with a data set which is effectively free of external backgrounds.

\bibliography{bibliography.bib}
\bibliographystyle{JHEP}

\end{document}